\documentclass[twocolumn]{aastex63}
\usepackage{epsfig}
\usepackage{natbib}
\usepackage{amsmath}
\usepackage{verbatim}
\usepackage{longtable}
\usepackage{hyperref}
\usepackage{supertabular}
\definecolor{blue}{RGB}{50, 80, 255}
\bibpunct{(}{)}{;}{a}{}{,}      
\newcommand{\nlines}{94}

\shorttitle{Spectroscopic Study of G238-44}
\shortauthors{T. Johnson et al}

\begin{document}

\title{Unusual Abundances from Planetary System Material Polluting the White Dwarf G238-44}

\author[0000-0002-1570-2203]{Ted M Johnson}
\affiliation{Department of Physics and Astronomy, University of California, Los Angeles, CA 90095-1562, USA}

\correspondingauthor{Ted Johnson}
\email{tedjohnson12@g.ucla.edu}

\author[0000-0001-5854-675X]{Beth L. Klein}
\affiliation{Department of Physics and Astronomy, University of California, Los Angeles, CA 90095-1562, USA}

\author[0000-0002-6164-6978]{D. Koester}
\affiliation{Institut fur Theoretische Physik und Astrophysik, University of Kiel, 24098 Kiel, Germany}

\author[0000-0001-9834-7579]{Carl Melis}
\affiliation{Center for Astrophysics and Space Sciences, University of California, San Diego, CA 92093-0424, USA}

\author[0000-0001-6809-3045]{B. Zuckerman}
\affiliation{Department of Physics and Astronomy, University of California, Los Angeles, CA 90095-1562, USA}

\author{M. Jura}\altaffiliation{deceased}
\affiliation{Department of Physics and Astronomy, University of California, Los Angeles, CA 90095-1562, USA}

\begin{abstract}
Ultraviolet and optical spectra of the hydrogen-dominated atmosphere
white dwarf star G238-44 obtained with FUSE, Keck/HIRES, HST/COS, and
HST/STIS reveal ten elements heavier than helium: C, N, O, Mg, Al, Si,
P, S, Ca, and Fe. G238-44 is only the third white dwarf with nitrogen
detected in its atmosphere from polluting planetary system material.
Keck/HIRES data taken on eleven nights over 24 years show no evidence for
variation in the equivalent width of measured absorption lines,
suggesting stable and continuous accretion from a circumstellar
reservoir. From measured abundances and limits on other elements we find
an anomalous abundance pattern and evidence for the presence of metallic
iron. If the pollution is from a single parent body, then it would have
no known counterpart within the solar system. If we allow for two
distinct parent bodies, then we can reproduce the observed abundances
with a mix of iron-rich Mercury-like material and an analog of an icy
Kuiper Belt object with a respective mass ratio of 1.7:1. Such
compositionally disparate objects would provide chemical evidence for
both rocky and icy bodies in an exoplanetary system and would be
indicative of a planetary system so strongly perturbed that G238-44 is
able to capture both asteroid- and Kuiper Belt-analog bodies
near-simultaneously within its $<$100 Myr cooling age.

\end{abstract}


\section{Introduction \label{sec:intro}}

It is now well established that single white dwarf (WD) stars with $T_{\rm eff}$ $\lesssim$ 25,000\,K whose spectra show absorption features from elements heavier than helium (high-Z) have accreted material from their extant planetary systems \citep[e.g.,][and references therein]{2014AREPS..42...45J, 2016NewAR..71....9F, 2018haex.bookE..14Z, 2021orel.bookE...1V}.  Perturbed planet(esimal)s in these evolved planetary systems \citep{2002ApJ...572..556D} will be tidally disrupted if they venture within the WD Roche radius, forming an accretion disk which ``pollutes'' the otherwise pure hydrogen and/or helium atmosphere of the WD star \citep{2003ApJ...584L..91J}.  Measuring abundances of high-Z elements in these WD atmospheres has proved to be a unique and powerful method to determine the compositions of exoplanetary bodies to a high degree of precision and sensitivity.

For the most part, parent body compositions have been found to be consistent with dry, volatile-poor rocky material typical of the inner solar system (e.g., \citealt{2006ApJ...653..613J}; \citealt{juraxu12}; \citealt{wilson16}; \citealt{2019Sci...366..356D}). However, examples exist of objects very rich in water-ice \citep[e.g.][]{2013Sci...342..218F, 2015MNRAS.450.2083R, 2019Natur.576...61G, 2021ApJ...914...61K, 2021ApJ...907L..35D} $-$ including an exo-Kuiper Belt analog \citep{2017ApJ...836L...7X} $-$ and bodies massive enough to have experienced igneous differentiation into cores  \citep{2011ApJ...732...90M, 2012MNRAS.424..333G} and crusts \citep{2011ApJ...739..101Z, 2017ApJ...834....1M}. Even with this assortment of measured exoplanetary compositions, it has consistently been the case that polluted white dwarf abundances can be reasonably fit with the compositions of known solar system bodies.


Theoretical works allow for pollution from multiple parent bodies \citep[e.g.,][]{2008AJ....135.1785J,2020MNRAS.491.4672T}, but to date all previously published polluted white dwarf systems are understood as accreting a single parent body. As we report herein, G238-44 appears to break the mold of past polluted white dwarf results. Either it is accreting material that is so alien as to have no known counterpart within the solar system, or it is simultaneously accreting from two compositionally distinct parent bodies.


G238-44 (13:38:50.47 +70:17:07.64; commonly used alternate names are EG\,102 and WD1337+705) is a warm (effective temperature $\approx$20,000\,K), hydrogen-dominated and metal-polluted atmosphere WD (spectral type DAZ).  G238-44 was identified almost 80 years ago by \citet{1942ApJ....96..315K} when there were fewer than 80 WDs known \citep{1945Sci...101...79L}. The proximity of G238-44 to Earth leads it to be quite bright for a WD at $\approx$12$^{\rm th}$\,mag and, combined with its relatively simple spectrum dominated by hydrogen Balmer lines (Figure \ref{fig:sed_b}), has seen the star extensively observed as both a spectrophotometric calibrator and a science object.

Absorption features of Si, C, and Al from International Ultraviolet Explorer (IUE) spectra provided the first evidence for elements other than H in the atmosphere of G238-44. This prompted debate
about whether the ground state and weakly excited ultraviolet (UV) lines were photospheric, intersteller, or circumstellar in origin \citep{1991ASIC..336..235V, 1997ApJ...474L.127H, 1998ApJS..119..207H}.  
On the other hand, Mg II 4481\,{\AA} $-$ arising from an excited state of 8.86 eV $-$ is always photospheric, and its detection in G238-44 \citep{1997ApJ...474L.127H} was the first observation of Mg in a DA of this temperature class.  Radiative levitation can support high-Z elements (particularly Si, C, and Al) against gravitational settling in WD atmospheres, but the effect was shown to be negligible for Mg at $T_{\rm eff} $ $\sim$20,000\,K \citep{1995ApJ...454..429C}.  Since the settling timescales for high-Z elements are on the order of $\sim$days for G238-44, the Mg detection was early evidence that observed atmospheric pollution in this class of WDs must be associated with current and ongoing accretion.  

Spitzer Infrared Array Camera 4.5$\micron$ and 7.9$\micron$ photometry of G238-44 shows no infrared excess from any putative circumstellar disk at those wavelengths \citep{2007ApJS..171..206M}.  \citet{2009ApJ...694..805F} report the detection of a faint MIPS 24$\micron$ source near G238-44's location, which would suggest a subtle excess from cool ($T$ $<$ 400K) dust if it is actually associated with the WD.  However, the authors conclude that the potential for source confusion is high, $\sim$25-70\%, and that the 24$\micron$ flux may not originate from G238-44.  Thus, despite ongoing accretion, the debris reservoir around this star is among the vast majority that go undetected \citep{2008AJ....135.1785J, 2015MNRAS.449..574R, 2019MNRAS.487..133W}. 





Optical detection of Ca was reported by \citet{1998ApJ...505L.143Z} and \citet{2003ApJ...596..477Z} as part of a larger study of WD pollution which concluded that neither accretion of material from the interstellar medium (ISM) nor exo-comets can easily explain the observations.  Subsequently, a suite of elements $-$ C, N, O, Si, S, and Fe $-$ detected in spectra from the Far Ultraviolet Spectroscopic Explorer (FUSE) were analyzed in the context of radiative levitation by \citet{2007ASPC..372..261D, 2010AIPC.1273..412D}. They concluded that ongoing accretion, rather than radiative support of material from a past pollution event, must be responsible for the observed atmospheric pollution in this warm DAZ. 

G238-44 earns the distinction of being one of just three known WDs to be polluted by planetary system material that produces detectable N absorption; 
the other two are G200-39 \citep{2017ApJ...836L...7X} and GD 378 \citep{2021ApJ...914...61K}. The detection of N is significant because it is typically indicative of exoplanetary ices or primitive volatile-rich material such as giant planets, Kuiper Belt-analog objects, and comet-like bodies.  



In this paper we collate data from a variety of UV and optical spectrographs (Section \ref{sec:obs}) 
to identify, characterize, and measure \nlines\, photospheric absorption lines in the spectrum of G238-44 (Section \ref{sec:measurement}). Section \ref{sec:parent} derives accretion rates for ten high-Z elements and demonstrates that these elements are present in ratios unlike any known single solar system body. We put forward a model for pollution by two compositionally distinct parent bodies and show it provides a reasonable fit to the available data.


\section{Observations \label{sec:obs}}
High resolution spectra of G238-44 were obtained from various space telescope archives and our own observations with Keck/HIRES. An observing log appears in Table \ref{tab:obs}.
Through this complementary combination of facilities and instruments we cover a wide range of wavelengths and are able to compile the suite of elements presented in this study.

\begin{table*}[ht]
\caption{Log of Observations \label{tab:obs}}
\begin{center}
\begin{tabular}{lcccccc}
\hline 
\hline
UT Date  & Instrument & $\lambda$ range  & exposure & R & Program ID & P.I. \\
		&		  & 		({\AA}) 	   &	(seconds) & & &  \\
\hline
1997 Jul 07 & HIRES Blue & 3760$-$5250    &  1800 & 34,000 & C08H & N. Reid \\
1998 Jan 23 & HIRES Blue & 3760$-$5250    &  900& 34,000 & U04H & B. Zuckerman \\
1999 Apr 20 & HIRES Blue & 3760$-$5250    &  2400& 34,000 & U06H & B. Zuckerman \\
1999 Jul 02 & HIRES Red & 4180$-$6710    &  3600& 25,000 & C04H & N. Reid \\
1999 Jul 03 & HIRES Red & 4180$-$6710    &  900& 25,000 & C04H & N. Reid \\
2001 May 05 & FUSE	             & 905$-$1180 & 10746& 15,000 & B119 & J. Holberg\\
2002 Jan 27 & FUSE               & 905$-$1180  & 17410& 15,000 & S601 & S. Friedman \\
2006 Jun 11 & HIRES Red & 5560$-$10160 & 	7200& 37,000  & U106Hr & B. Hansen \\
2006 Jun 17 & HIRES Blue & 3050$-$5940    &  3600& 37,000 & U106Hr & B. Hansen \\
2007 May 05 & HIRES Blue & 3050$-$5940   &  1800& 37,000 & U103Hb & B. Hansen\\
2007 May 06 & HIRES Blue & 3050$-$5940   &  3600& 37,000 & U103Hb & B. Hansen\\
2009 Dec 26 & STIS E230H         & 2577$-$2835 & 1200& 114,000 & 11568 & S. Redfield \\
2009$-$2021 & COS G185M      & 1730$-$1970 & 770& 20,000 & \multicolumn{2}{c}{{\it Various}} \\
2013$-$2017 & COS G160M      & 1430$-$1800 & 301& 20,000 & {\it Various} & S. Penton \\
2016 Jun 03 & HIRES Blue & 3150$-$5940   &  1800& 37,000 & N171Hb & S. Redfield\\
2021 Dec 13 & HIRES Blue & 3050$-$5940   &  2000& 37,000 & U128Hb & C. Melis \\
\hline
\multicolumn{7}{p{14cm}}{Note. --   COS spectra are from many observations combined. See Section \ref{subsec:cos}. 
}
\end{tabular}
\end{center}
\end{table*}

\subsection{FUSE \label{subsec:fuse}}
G238-44 was observed in TTAG mode with the LWRS aperture of FUSE \citep{2000ApJ...538L...1M} on May 05, 2001 (PI J. Holberg) and again on January 27, 2002 (PI S. Friedman).  While the 
wavelength range extended down to 905\,\r{A}, the flux for G238-44 is essentially zero for wavelengths shorter than 955\,\r{A}.  These data were first published by \citet{2003A&A...405.1153H} who examined Lyman series quasi-molecular satellite features, and were revisited by \citet{2007ASPC..372..261D, 2010AIPC.1273..412D} and \citet{2014MNRAS.440.1607B} in the context of high-Z abundances and radiative levitation.

Each observation included the four optical path channels (LiF1, LiF2, SiC1, SiC2) focused on two detector segments (A, B), which resulted in eight independent spectra with varying degrees of overlap in wavelength coverage.  For most of our analysis we use a fully combined total spectrum provided by K.\ Long and B.\ G\"{a}nsicke (2012, private communication).  
We found that some unusual features appear in the region of the O\,I $\lambda$1152 line, which is covered by two of the eight channels/segments (LiF1B and LiF2A). To investigate the possible cause, we downloaded the uncombined data products available through the Mikulski Archive for Space Telescopes (MAST). We inspected the individual spectra as well as the nighttime-only versions of both epochs of observations, finding that the O\,I $\lambda$1152 feature has contamination/distortions in the LiF2A spectra from both epochs (also in the night-only data) while the LiF1B spectra from both epochs look ``normal''. We thus created an uncontaminated version of the 1100-1180\,{\AA} region by selecting only the LiF1B segments from both epochs.

We also found the multiplet of N\,II lines around 1084\,\r{A} to be unusually broad in the fully combined spectrum.  Inspection of the individual segments revealed that the N\,II lines in the SiC1A segment of the S601 exposure were distorted (broader and shallower) compared to their appearance in the other three segments (SiC1A and SiC2B of B119 and SiC2B of S601).  Thus, for the region of 1080-1090\,{\AA}, we created a combined spectrum omitting the S601-SiC1A segment. Since this was done outside of the CalFUSE pipeline, we do not report radial velocities for these lines.



The majority of detected absorption lines are found in these FUSE spectra, in which we identify transitions from C, N, O, Si, P, S, and Fe.

\begin{figure*}[]
\begin{center}
\includegraphics[width=140mm]{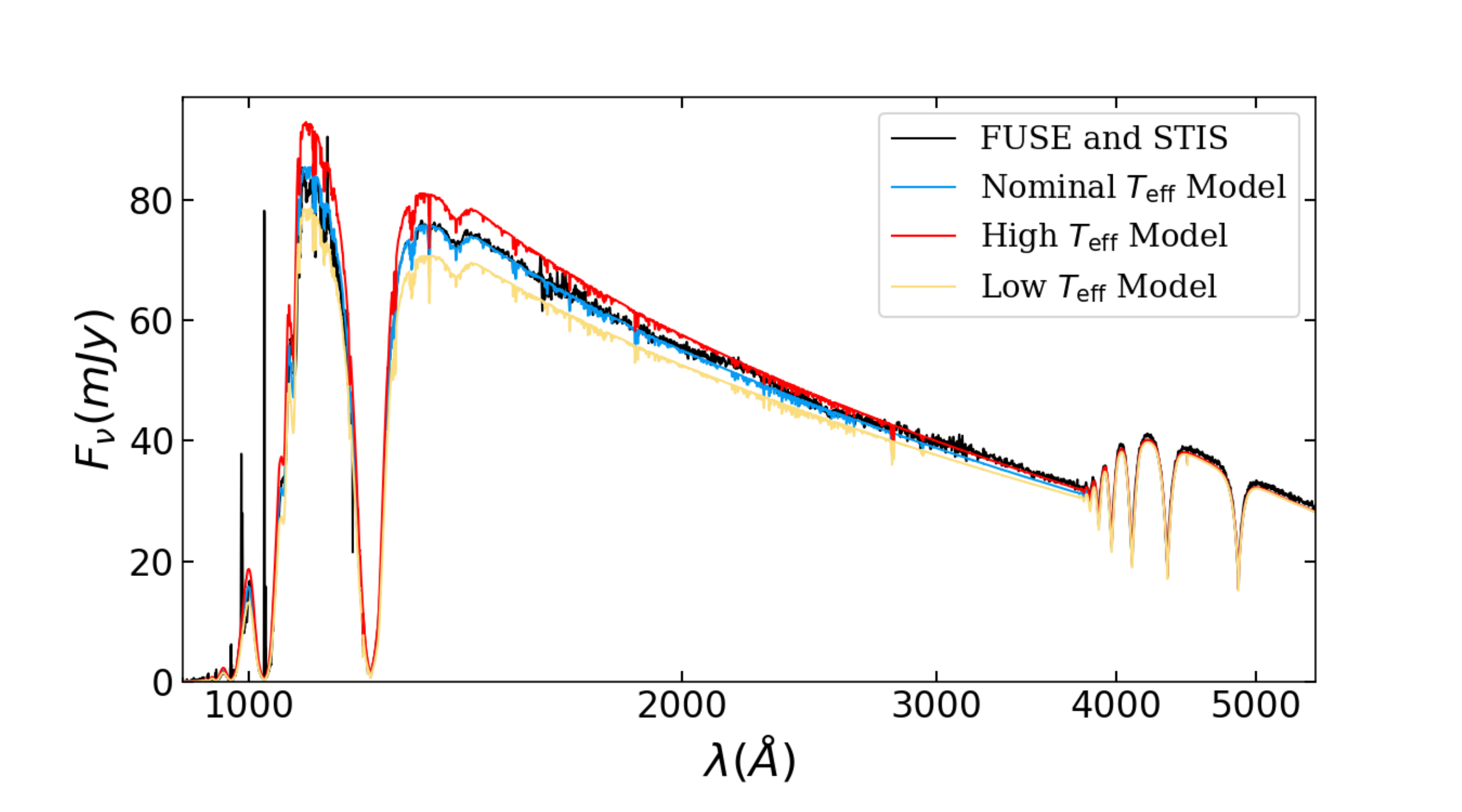}
\caption{Ultraviolet and blue-optical spectra of G238-44 and models as described in Section \ref{subsec:model}. Low resolution and high SNR STIS spectra are displayed for wavelengths longer than
1300\,\r{A} in preference to lower SNR COS spectra; STIS spectra are used as obtained from MAST without further processing.
The model produced with nominal values of $T_{\rm eff}$ and log$\,g$ {is an excellent match to the flux calibrated spectra.} 
}
\label{fig:sed_b}
\end{center}
\end{figure*}

\subsection{HIRES \label{subsec:hires}}
Throughout the timespan of 1997-2021 G238-44 was observed on 11 nights with the High Resolution Echelle Spectrograph \citep[HIRES,][]{1994SPIE.2198..362V} at the W.\ M.\ Keck Observatory (see Table \ref{tab:obs}). All observations reported used the C5 decker (mapping to a $\approx$1.15$'$$'$ slit)
and used either the red- or blue-optimized collimator.
The 1997-1999 HIRES data were acquired with the pre-2004 single-CCD detector setup
and were published by \citet{1998ApJ...505L.143Z} and \citet{2003ApJ...596..477Z}.
All subsequent data were taken using the updated three-detector mosaic, which resulted in improved resolution, wavelength coverage, and signal-to-noise ratio (SNR) compared to the older spectra. 


Data were reduced with either PyRAF \citep{2012ascl.soft07011S} or the MaunaKea Echelle Extraction (MAKEE\footnote{\url{https://www.astro.caltech.edu/~tb/makee/}}) pipeline and subsequently processed with methods following \citet{2010ApJ...709..950K}.
The spectra were continuum normalized by fitting low order cubic spline functions and combined using PyRAF. Relatively long integrations obtained in 2006-2007 resulted in an exceedingly high quality spectrum with SNR per pixel  
of 190 near 3200\,\r{A}, 370 near 3930\,\r{A}, 220 near 4500\,\r{A}, and 260 near 6350\,\r{A}. The high resolution and high SNR of these data enable the detection of lines with equivalent widths on the order of a few m\r{A} and allow us to  
report new optical detections of Si and Fe (previously seen Mg and Ca are also recovered).

\subsection{STIS \label{subsec:stis}}
G238-44 was observed by the Space Telescope Imaging Spectrograph\footnote{\url{https://hst-docs.stsci.edu/stisihb}} (STIS) aboard the Hubble Space Telescope (HST) on December 26, 2009 using the 0.2X0.2 aperture and E230H grating. 
The resulting spectrum has a SNR per pixel of 5.
These data were published by \citet{2014ApJ...787...75M} in an analysis using ground state near-UV Mg and Fe lines to study the local interstellar medium.  In our re-analysis of this spectrum we find
that the origin of the Mg and Fe lines is almost entirely, if not completely, due to photospheric absorption (see Sections \ref{subsubsec:Mg} and \ref{subsubsec:Fe}).


\subsection{COS \label{subsec:cos}}
These HST data were taken for the purpose of calibrating wavelength sensitivity and offsets.  Between November 1, 2013 and September 6, 2017 G238-44 was observed 13 times with the HST  Cosmic Origins Spectrograph\footnote{\url{https://hst-docs.stsci.edu/cosihb}} (COS) using the Primary Science Aperture (PSA) and G160M grating (Program IDs 13124, 13526, 13972, 14440, 14857). When combined, the spectra provide a SNR per resolution element of 15. 

G238-44 was also observed by COS using PSA and the G185M grating 11 times between July 21, 2009 and January 02, 2021 (Program IDs 11479, PI A. Aloisi; 13124, 13526, 13972, 14440, 14857, 15389, PI S. Penton; 15542, 15780, PI D. Sahnow). Each of the 11 exposures has similar spectral resolution and SNR, but cover different wavelength regions. 
The spectra were combined to increase SNR in overlapping spectral regions, resulting
in SNR per resolution element around 1862\,\r{A} of $\approx$20. 

We identify absorption lines from Al II and Al III in the combined COS spectra.

\section{Models and Measurements \label{sec:measurement}}
\subsection{Stellar Parameters\label{subsec:model}}
White dwarf model atmospheres and synthetic spectra used in this work are described in \citet{2010MmSAI..81..921K}. We fit a pure hydrogen atmosphere model to the G238-44 Gaia {DR2} photometry and parallax to obtain the atmospheric parameters listed in Table \ref{tab:params}. {The initial fit suggested a $T_{\rm eff}$ error of 101 K and resulted in a reduced $\chi ^2$ of 22. This fit utilized unrealistically small magnitude errors from the Gaia DR2 catalog. Thus} errors on the photometry were scaled up by a multiplicative factor until we obtained a final best-fit $\chi^2$$\sim$1.

Figure \ref{fig:sed_b} shows flux-calibrated UV and optical spectra overlaid with our models at the lower, nominal, and upper limits of $T_{\rm eff}$/log\,$g$ solutions.  The excellent {match} of the nominal model to the flux-calibrated spectra adds confidence that these stellar parameters {indeed provide an accurate description of the available data for G238-44}, especially given that at a distance of only 26.5\,pc we do not expect reddening to affect the UV fluxes.
We note that GALEX FUV and NUV photometric measurements of G238-44 are available, but its brightness puts it in the non-linear regime of the instrument where the GALEX magnitudes are clearly affected according to the known linearity response plot (\url{https://asd.gsfc.nasa.gov/archive/galex/Documents/brightness_limits.html}).

\begin{table}
\caption{Summary of Parameters \label{tab:params}}
\begin{center}
\begin{tabular}{lrrl}
\hline 
\hline													
$G$ (mag)  &  12.78 (0.01)  \\
Distance (pc)  & 26.5 (0.2) \\
$T_{\rm eff}$ (K)  & 20546 (474) \\
log $g$  & 7.95 (0.03)  \\
$M_{\rm WD} (M_\sun$)  &   0.597 (0.018) \\   
$R_{\rm WD} (R_\sun$)  &   0.0136 (0.0003)  \\    
Cooling age (Myr) & 50 (2)  \\    
$\dot{M}$ (g s$^{-1}$)   &  5.8 x 10$^7$   \\
Grav.~redshift (km s$^{-1}$)  &  27.9 (1.5) \\  
V$_{\rm r,helio}$ (km s$^{-1}$) & 1 (2)  \\
V$_{\rm systemic}$ (km s$^{-1}$) & $-$27 (3)  \\
\hline
\end{tabular}
\end{center}
\tablecomments{$G$ magnitude and distance are from Gaia DR2 and EDR3.  $M_{\rm WD}$, $R_{\rm WD}$, gravitational redshift, and cooling age are from the Montreal White Dwarf Database \citep[MWDD;][]{2017ASPC..509....3D}\footnote{\url{http://dev.montrealwhitedwarfdatabase.org/evolution.html}}. Uncertainties given in parentheses represent the range in values for each parameter considered at the upper and lower limits of the $T_{\rm eff}$/\,log\,$g$ models.
$\dot{M}$ is the total mass flow rate (see Section \ref{sec:parent}). }
\end{table}

\subsection{Metal Absorption Lines \label{subsec:absorp_lines}}
Table \ref{tab:linelist} presents a linelist for the $\nlines$ absorption lines from heavy elements
we identified in the collected spectroscopic datasets.
Each line was matched to an atomic transition using the Vienna Atomic Line Database (VALD\footnote{\url{http://vald.astro.uu.se/}}) and/or the National Institute of Standards and Technology (NIST\footnote{\url{https://www.nist.gov/pml/atomic-spectra-database}}) Atomic Spectra Database.  Both the lower energy level of the transition and the log($gf$) were considered when determining the line identifications and their inclusion in the line list.
We list the adopted laboratory wavelengths, lower energy levels, and oscillator strengths used for modeling lines in Table \ref{tab:linelist}.

\begin{figure}[]
\includegraphics[width=8cm]{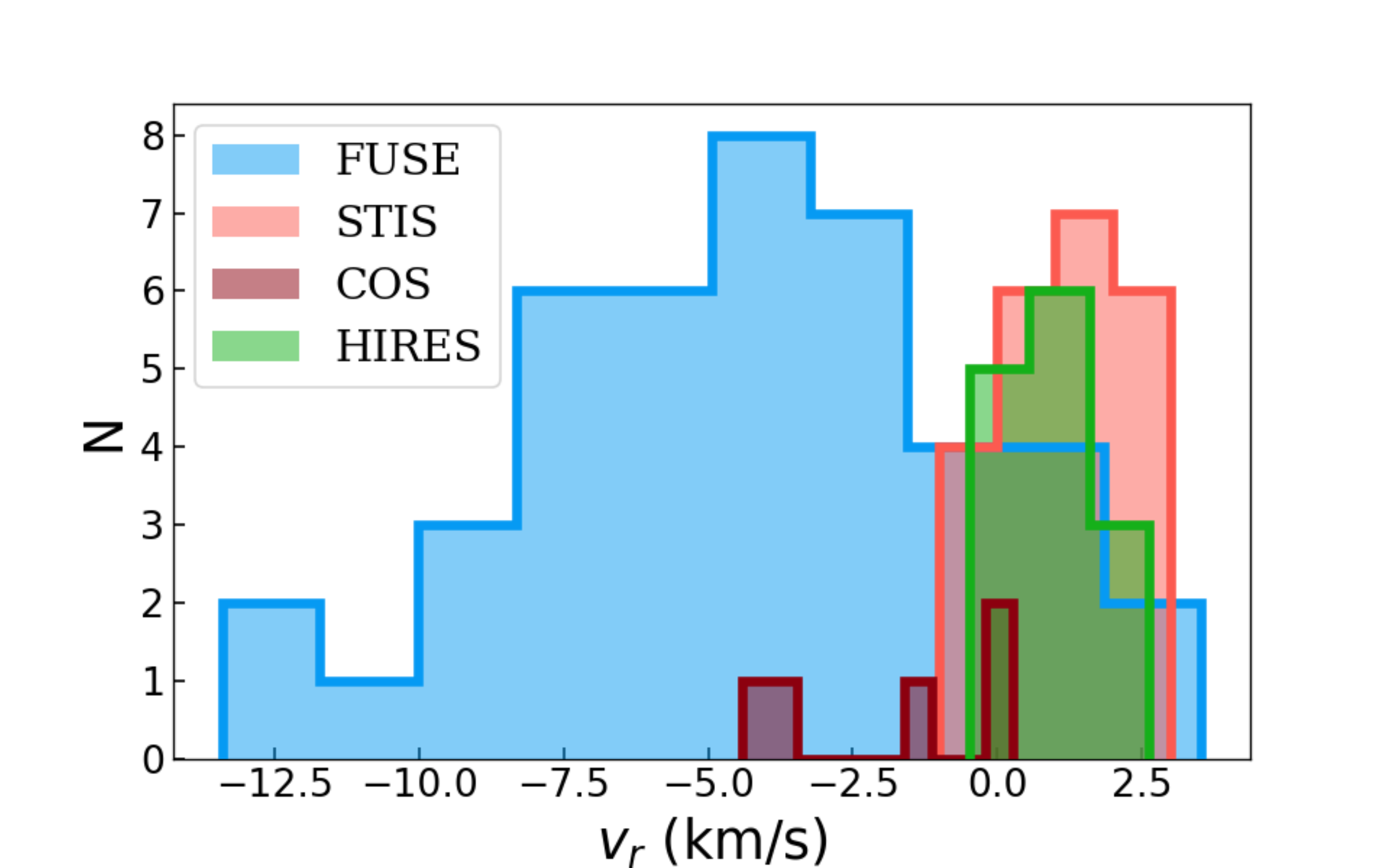}
\caption{Distribution of heliocentric radial velocities measured for the absorption lines detected in spectra of G238-44. 
See also Section \ref{subsec:absorp_lines}. }
\label{fig:hist}
\end{figure}

Equivalent widths (EWs) were measured by fitting Voigt profiles to absorption lines with IRAF's {\sf splot} routine. Uncertainties for each line were calculated by summation in quadrature of {\sf splot}'s line-fitting uncertainties and the RMS of (typically) three EW measurements where the continuum level was varied within the noise between each measurement.   Line centers were measured either from the {\sf splot} profile fitting or visual inspection of the line, from which we calculated heliocentric radial velocities reported in Table \ref{tab:linelist} and displayed in Figure \ref{fig:hist}. In the case of a blended feature, e.g.\ Mg~II 4481 \r{A}, we use the sum of EWs resulting from fitting multiple profiles simultaneously with {\sf splot}'s `deblending' routine and do not report radial velocity measurements.

While the measurements from the different instruments are in general agreement, the peak of the distribution of FUSE RVs is offset from HIRES and STIS by about -5 km s$^{-1}$, and it has a wider spread.  This is consistent with the known absolute and relative wavelength calibration uncertainties of FUSE spectra (see FUSE Data Handbook\footnote{\url{https://archive.stsci.edu/fuse/DH_Final/index.html}}). 
From the 14 high-Z lines in the HIRES data we find an average RV of 0.9\,km\,s$^{-1}$ with a standard deviation of 0.9\,km\,s$^{-1}$.  These are in agreement with H$\alpha$ and H$\beta$, both of which have RVs of 1 km/s.  Adding in an absolute wavelength scale calibration uncertainty of $\simeq$1\,km\,s$^{-1}$, we find a heliocentric RV (which includes the gravitational redshift component) for G238-44 of 1 $\pm$ 2\,km\,s$^{-1}$.  




\startlongtable
\begin{deluxetable}{lccccc}
\tabletypesize{\scriptsize}
\tablecolumns{6}
\tablecaption{Absorption Line List\label{tab:linelist}}
\tablehead{\colhead{$\lambda_{\rm lab}$} & \colhead{Species}  & \colhead{$V_{\rm r}$}  &  \colhead{EW} & \colhead{$\chi$} & \colhead{log($gf$)}  \\
\colhead{ ({\AA}) } & \colhead{} & \colhead{(km/s)}  &  \colhead{ (m{\AA})} & \colhead{ (eV)} & \colhead{} 
}
\startdata
1009.858  & C II & 2.7 &   16.7  $\pm$  5.0 &5.332 & -0.462  \\ [-1mm] 
1010.083  & C II & 0.9 &   25.9  $\pm$  7.2 &5.334 & -0.161  \\ [-1mm] 
1010.371  & C II & 1.0 &   24.1  $\pm$  5.5 &5.338 & 0.015  \\ [-1mm] 
1036.337  & C II & -3.0 &   95.8  $\pm$  22.8 &0.000 & -0.624  \\ [-1mm] 
1037.018  & C II & -3.6 &   110.7  $\pm$  19.7 &0.008 & -0.328  \\ [-1mm] 
1065.891  & C II & -5.6 &  28.9  $\pm$  3.2 &9.290 & -0.052  \\ [-1mm] 
1066.133  & C II & 0.8 &   25.3  $\pm$  3.5 &9.290 & -0.307  \\ [-1mm] 
1174.933  & C III & -0.8 &  24.2  $\pm$  5.1 &6.496 & -0.468  \\ [-1mm] 
1175.263  & C III & -2.5 &   25.6  $\pm$  5.4 &6.493 & -0.565  \\ [-1mm] 
1175.6 	& C III &  &   85.1  $\pm$  14.3 &6.496 & -0.690  \\ [-1.5mm] 
& & & & 6.503 & 0.009  \\ [-1mm] 
1175.987  & C III & 0.8 &    16.5  $\pm$  4.5 &6.496 & -0.565  \\ [-1mm] 
1176.370  & C III & 0.0 &   16.2  $\pm$  4.7 &6.503 & -0.468  \\ [-1mm] 
1083.990  & N II & 0.0 &    47.5  $\pm$  4.5 &0.000 & -0.939  \\ [-1mm] 
1084.580  & N II & 0.0 &   60.8  $\pm$  4.3 &0.006 & -0.587  \\ [-1mm] 
1085.546  & N II & 0.0 &   40.2  $\pm$  4.7 &0.016 & -1.078  \\ [-1mm] 
1085.700  & N II & 0.0 &   55.7  $\pm$  4.4 &0.016 & -0.320  \\ [-1mm] 
1152.150  & O I & 0.0 &   30.4  $\pm$  6.0 &1.967 & -0.268  \\ [-1mm] 
2791.600  & Mg II & 2.4 &  81.2  $\pm$  25.9 &4.422 & 0.280  \\ [-1mm] 
2796.352  & Mg II & 2.3 &  226.3  $\pm$  38.1 &0.000 & 0.100  \\ [-1mm] 
2798.823  & Mg II & 0.0 &   95.8  $\pm$  21.4 &4.434 & 0.530  \\ [-1mm] 
2803.531  & Mg II & 1.5 &  171.0  $\pm$  21.6 &0.000 & -0.210  \\ [-1mm] 
4481		& Mg II &  &   95.3  $\pm$  23.1 &8.864 & 0.740  \\ [-1.5mm] 
& & & & 8.864 & -0.560  \\ [-1.5mm] 
& & & & 8.864 & 0.590  \\ [-1mm] 
1670.787  & Al II & -3.5 &   82.0  $\pm$  10.9 &0.000 & 0.263  \\ [-1mm] 
1721  & Al II &  &  25.7  $\pm$  7.1 &4.644 & -0.170  \\ [-1.5mm] 
& & & & 4.644 & 0.307  \\ [-1mm] 
1724.982  & Al II & -4.4 &   73.4  $\pm$  12.3 &4.659 & 0.577  \\ [-1mm] 
1854.716  & Al III & 0.0 &    123.9  $\pm$  19.3 &0.000 & 0.060  \\ [-1mm] 
1862.790  & Al III & 0.3 &    87.3  $\pm$  21.3 &0.000 & -0.240  \\ [-1mm] 
992.683  & Si II & 3.5 &   75.0  $\pm$  17.6 &0.036 & -0.277  \\ [-1mm] 
3856.019  & Si II & 0.6 &    1.8  $\pm$  0.4 &6.859 & -0.406  \\ [-1mm] 
3862.595  & Si II & 0.4 &   1.3  $\pm$  0.4 &6.858 & -0.860  \\ [-1mm] 
4128.054  & Si II & 1.2 &  1.7  $\pm$  0.8 &9.837 & 0.359  \\ [-1mm] 
4130.894  & Si II & -0.5 &  1.5  $\pm$  0.5 &9.839 & 0.552  \\ [-1mm] 
6347.109  & Si II & 0.6 &  10.8  $\pm$  1.5 &8.121 & 0.297  \\ [-1mm] 
6371.370  & Si II & 1.5 &   10.3  $\pm$  2.4 &8.121 & -0.003  \\ [-1mm] 
1108.358  & Si III & -5.7 &   67.3  $\pm$  3.5 &6.537 & -0.056  \\ [-1mm] 
1110  & Si III &  &   116.9  $\pm$  3.3 &6.553 & -0.186  \\ [-1.5mm] 
& & & & 6.553 & 0.294  \\ [-1mm] 
1113  & Si III &  &  158.9  $\pm$  5.4 &6.585 & -1.356  \\ [-1.5mm] 
& & & & 6.585 & -0.186  \\ [-1.5mm] 
& & & & 6.585 & 0.564  \\ [-1mm] 
1141.579  & Si III & -0.8 &   14.3  $\pm$  3.0 &16.098 & 0.470  \\ [-1mm] 
1144.309  & Si III & -8.1 &   18.4  $\pm$  3.4 &16.131 & 0.740  \\ [-1mm] 
1128.340  & Si IV & -3.2 &   12.2  $\pm$  1.9 &8.896 & 0.470  \\ [-1mm] 
1153.995  & P II & -12.2 &    9.9  $\pm$  3.6 &0.058 & -0.032  \\ [-1mm] 
1124.395  & S II & -8.4 &   4.7  $\pm$  2.4 &3.041 & -1.190  \\ [-1mm] 
1124.986  & S II & -4.6 &   18.8  $\pm$  3.8 &3.047 & -0.630  \\ [-1mm] 
1131.059  & S II & -4.0 &    12.7  $\pm$  4.5 &3.041 & -0.990  \\ [-1mm] 
1131.657  & S II & -5.6 &   3.4  $\pm$  1.1 &3.047 & -1.360  \\ [-1mm] 
3158.869  & Ca II & 2.0 &   8.5  $\pm$  0.9 &3.123 & 0.241  \\ [-1mm] 
3179.331  & Ca II & 1.6 &   4.3  $\pm$  1.2 &3.151 & 0.499  \\ [-1mm] 
3736.902  & Ca II & 2.6 &    0.9  $\pm$  0.3 &3.151 & -0.173  \\ [-1mm] 
3933.663  & Ca II & 0.5 &    28.5  $\pm$  0.5 &0.000 & 0.105  \\ [-1mm] 
3968.469  & Ca II & 0.3 &   5.5  $\pm$  0.6 &0.000 & -0.200  \\ [-1mm] 
3349.402  & Ti II &  & $<$  0.7    &0.049 & 0.530  \\ [-1mm] 
1004.670  & Ti III &  & $<$  7.2    &4.764 & 0.510  \\ [-1mm] 
3132.053  & Cr II &  & $<$  1.3   &2.483 & 0.423  \\ [-1mm] 
3441.987  & Mn II &  & $<$  0.3    &1.776 & -0.360  \\ [-1mm] 
1063.176  & Fe II & -3.9 &   11.5  $\pm$  5.0 &0.000 & -0.262  \\ [-1mm] 
1068.346  & Fe II & -3.6 &   20.9  $\pm$  3.4 &0.048 & -0.662  \\ [-1mm] 
1096.607  & Fe II & -8.2 &   8.5  $\pm$  2.9 &0.048 & -0.790  \\ [-1mm] 
1096.877  & Fe II & -13.4 &   9.8  $\pm$  3.2 &0.000 & -0.486  \\ [-1mm] 
1121.975  & Fe II & -6.2 &   11.5  $\pm$  2.1 &0.000 & -0.538  \\ [-1mm] 
1125.448  & Fe II & -8.9 &  10.9  $\pm$  4.0 &0.000 & -0.807  \\ [-1mm] 
1145  & Fe II &  & 25.2  $\pm$  3.1 &0.000 & 0.037  \\ [-1.5mm] 
& & & & 0.000 & -0.081  \\ [-1mm] 
1148.277  & Fe II & -10.4 &   19.4  $\pm$  3.7 &0.048 & -0.179  \\ [-1mm] 
1151.146  & Fe II & -7.4 &  10.1  $\pm$  3.4 &0.083 & -0.451  \\ [-1mm] 
2586.650  & Fe II & 2.0 &   22.7  $\pm$  6.6 &0.000 & -1.490  \\ [-1mm] 
2592.318  & Fe II & 2.7 &   28.9  $\pm$  8.7 &1.040 & -0.504  \\ [-1mm] 
2593.560  & Fe II & 1.0 &    29.3  $\pm$  8.9 &4.076 & 0.646  \\ [-1mm] 
2599.147  & Fe II & 1.0 &   36.8  $\pm$  7.6 &0.048 & -0.062  \\ [-1mm] 
2600.173  & Fe II & 0.9 &   57.8  $\pm$  11.5 &0.000 & 0.384  \\ [-1mm] 
2607.867  & Fe II & 1.3 &   35.7  $\pm$  7.9 &0.083 & -0.132  \\ [-1mm] 
2612.654  & Fe II & 2.1 &    39.5  $\pm$  10.4 &0.048 & 0.000  \\ [-1mm] 
2614.605  & Fe II & 0.0 &   50.8  $\pm$  10.9 &0.107 & -0.355  \\ [-1mm] 
2626.451  & Fe II & -0.3 &   30.5  $\pm$  10.4 &0.048 & -0.468  \\ [-1mm] 
2629.078  & Fe II & -0.6 &   37.0  $\pm$  7.9 &0.121 & -0.457  \\ [-1mm] 
2631.833  & Fe II & -0.6 &   31.5  $\pm$  9.8 &0.107 & -0.304  \\ [-1mm] 
2632.108  & Fe II & 2.1 &   26.4  $\pm$  7.9 &0.083 & -0.306  \\ [-1mm] 
2740.359  & Fe II & 2.7 &   51.4  $\pm$  13.7 &0.986 & 0.302  \\ [-1mm] 
2744.009  & Fe II & 0.8 &    25.2  $\pm$  7.3 &1.097 & -0.054  \\ [-1mm] 
2747.296  & Fe II & 1.2 &    44.6  $\pm$  12.8 &1.076 & 0.151  \\ [-1mm] 
2747.794  & Fe II & 0.9 &   40.5  $\pm$  14.4 &1.040 & 0.051  \\ [-1mm] 
2750.134  & Fe II & 1.4 &   53.8  $\pm$  14.7 &1.040 & 0.294  \\ [-1mm] 
2756.552  & Fe II & 1.0 &    58.8  $\pm$  13.1 &0.986 & 0.390  \\ [-1mm] 
2768.321  & Fe II & -0.2 &   25.0  $\pm$  9.2 &3.245 & 0.405  \\ [-1mm] 
3213.309  & Fe II & 1.5 &    2.0  $\pm$  0.3 &1.695 & -1.250  \\ [-1mm] 
3227.742  & Fe II & -0.2 &    3.6  $\pm$  0.5 &1.671 & -0.960  \\ [-1mm] 
5169.033  & Fe II & 0.6 &   3.0  $\pm$  1.2 &2.891 & -1.000  \\ [-1mm] 
1098.241  & Fe III & -1.1 &   11.7  $\pm$  2.7 &6.250 & -0.550  \\ [-1mm] 
1122.524  & Fe III & -6.1 &    52.4  $\pm$  3.1 &0.000 & -0.156  \\ [-1mm] 
1124.874  & Fe III & -5.1 &    38.3  $\pm$  4.9 &0.054 & -0.447  \\ [-1mm] 
1126.723  & Fe III & -3.3 &    29.5  $\pm$  6.9 &0.092 & -0.875  \\ [-1mm] 
1128.042  & Fe III & -2.7 &    23.5  $\pm$  2.2 &0.054 & -0.746  \\ [-1mm] 
1128.718  & Fe III & -4.5 &    33.1  $\pm$  3.2 &0.092 & -0.653  \\ [-1mm] 
1129.185  & Fe III & -2.4 &   23.2  $\pm$  2.8 &0.116 & -0.767  \\ [-1mm] 
1130.397  & Fe III & -4.0 &    16.7  $\pm$  2.3 &0.127 & -1.120  \\ [-1mm] 
1131.189  & Fe III & -1.9 &    19.3  $\pm$  7.0 &0.116 & -1.244  \\ [-1mm] 
1131.908  & Fe III & -2.4 &   9.5  $\pm$  2.5 &0.092 & -1.594  \\ [-1mm] 
1143.666  & Fe III & -9.7 &   8.3  $\pm$  5.6 &3.809 & -0.966  \\ [-1mm] 
1140.459  & Ni II &  & $<$  7.7   &1.254 & -0.494  \\ 
\enddata
\tablecomments{$\lambda < 3000$\,\r{A} given in vacuum, $\lambda \geq 3000$\,\r{A} given in air. The atomic data are the values used in our models for this analysis.  Features at 1175.6, 4481, 1721.3, 1110, 1113.2, and 1144.9 are blended lines and the quoted EW is the sum of the blend. }
\end{deluxetable}


\begin{figure*}[]
\fig{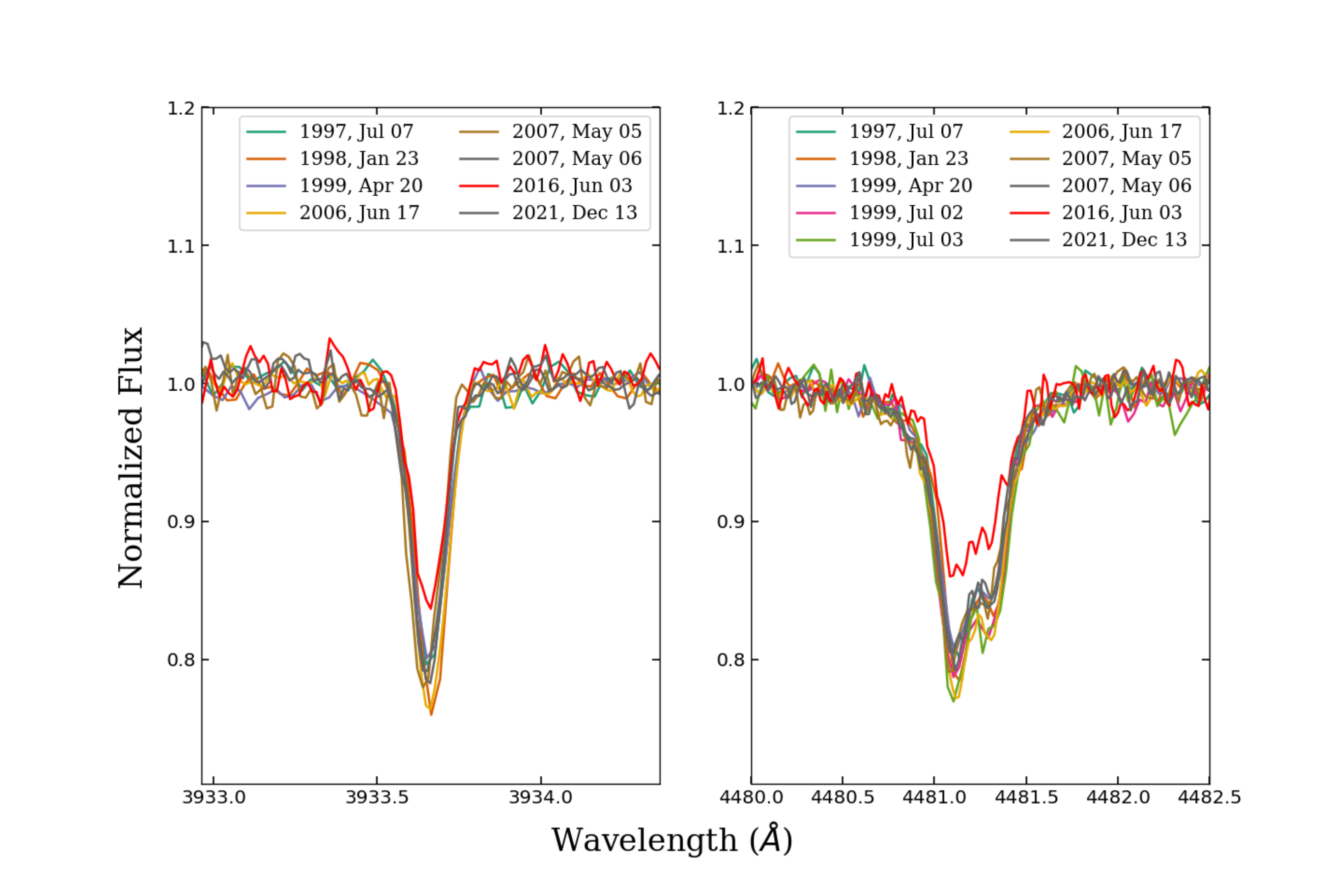}{5in}{}
\caption{Multiple epochs of HIRES spectra showing Ca~II\,K (left) and Mg~II\,$\lambda$4481 (right). With the exception of the anomalous 2016 epoch, the line strengths all look the same over the 24 year period during which the spectra were obtained.}
\label{fig:ew_change}
\end{figure*}

\begin{deluxetable}{lcc}
\tablecolumns{3}
\tabletypesize{\normalsize}
\tablecaption{HIRES Epoch EW Measurements \label{tab:epochs}   }
\tablehead{\colhead{Date} & \colhead{Ca K (m\r{A})} & \colhead{MgII $\lambda$4481 (m\r{A})} }
\startdata
1997 Jul 07 & $27.1 \pm 2.4$ & $95.0 \pm 8.7$ \\
1998 Jan 23 & $28.2 \pm 1.7$ & $99.1 \pm 7.4$ \\ 
1999 Apr 20 & $25.8 \pm 1.7$ & $95.7 \pm 10.8$\\
1999 Jul 02 & - & $104.0 \pm 6.9$ \\
1999 Jul 03 & - & $98.3 \pm 11.8$ \\
2006 Jun 17 & $29.9 \pm 2.1$ & $109.3 \pm 4.6$ \\
2007 May 05 & $29.3 \pm 4.8$ & $95.0 \pm 13.6$\\
2007 May 06 & $27.3 \pm 2.4$ & $100.9 \pm 4.7$ \\
2016 Jun 03 & $22.6 \pm 2.9$ & $59.1 \pm 12.3$ \\
2021 Dec 13 & $30.5 \pm 2.9$ & $93.5 \pm 7.1$ \\
\enddata
\tablecomments{Lines were measured in the reduced spectra prior to being normalized and combined.  Each measurement is an average of the multiple exposures and echelle orders from that epoch. }
\end{deluxetable}

Warm accreting WDs with short settling times would be expected to display variations in atmospheric pollution if there are variations in the accretion rate or inhomogeneities in the accretion-diffusion process, for example inefficient horizontal spreading of accreted material on the surface of the WD \citep{2021MNRAS.503.1646C}.  However, robust evidence for atmospheric abundance variability in such systems has not yet been found \citep{2008ApJ...677L..43D, 2018MNRAS.481.2601F, 2019MNRAS.483.2941W}.
With settling times on the order of days and being bright enough to produce high-SNR high-resolution spectra, G238-44 is a prime candidate to monitor for changes in high-Z absorption lines over time. 

Table \ref{tab:epochs} and Figure \ref{fig:ew_change} show the EWs and profiles of Ca~II\,K and Mg~II 4481 \r{A} in each epoch they have been observed with HIRES.  
With the exception of a single HIRES blue spectrum from 2016 (PI: S. Redfield), all observations
result in equivalent widths and line profiles that, within the uncertainties, are consistent from one epoch to another.
With just one inconsistent spectrum out of ten, we can only say that the 2016 measurement appears anomalous for some reason $-$ as it stands, we are currently unable to provide a definitive explanation. We do not include the 2016 spectrum in any analysis due to the unexplained low line EWs. A second epoch with a significant EW change would be required to convincingly argue that this WD has displayed robust and meaningful line strength variations. 

\begin{figure*}[]
\fig{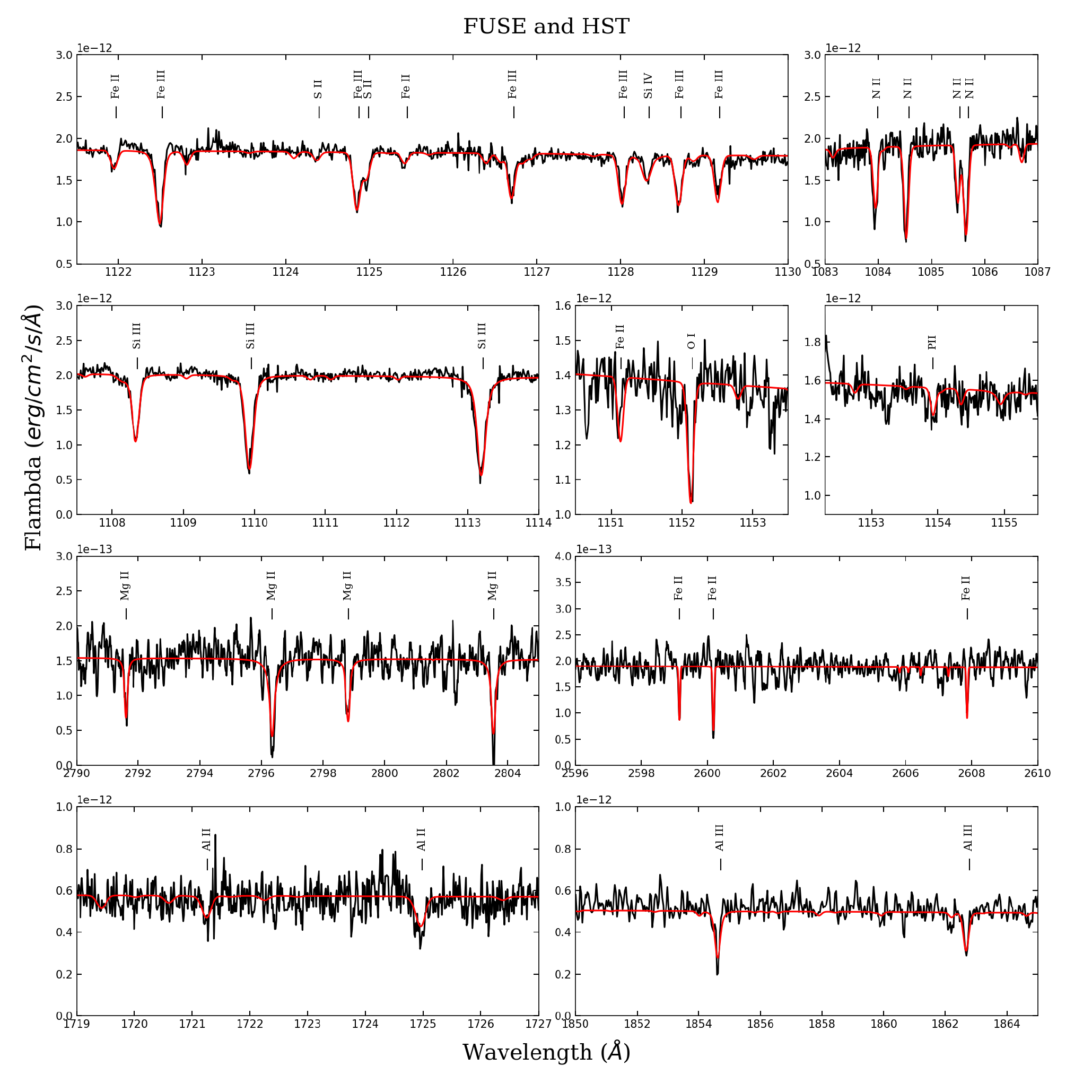}{6.5in}{}
\caption{Select regions of FUSE and HST spectra; the model with the best-fit abundances is overplotted in red. The STIS spectrum (showing the 2800\,\r{A} Mg\,II lines and 2600\,\r{A} Fe\,II lines) has been smoothed with a 5 pixel boxcar function.}
\label{fig:fuse}
\end{figure*}

\begin{figure*}[]
\fig{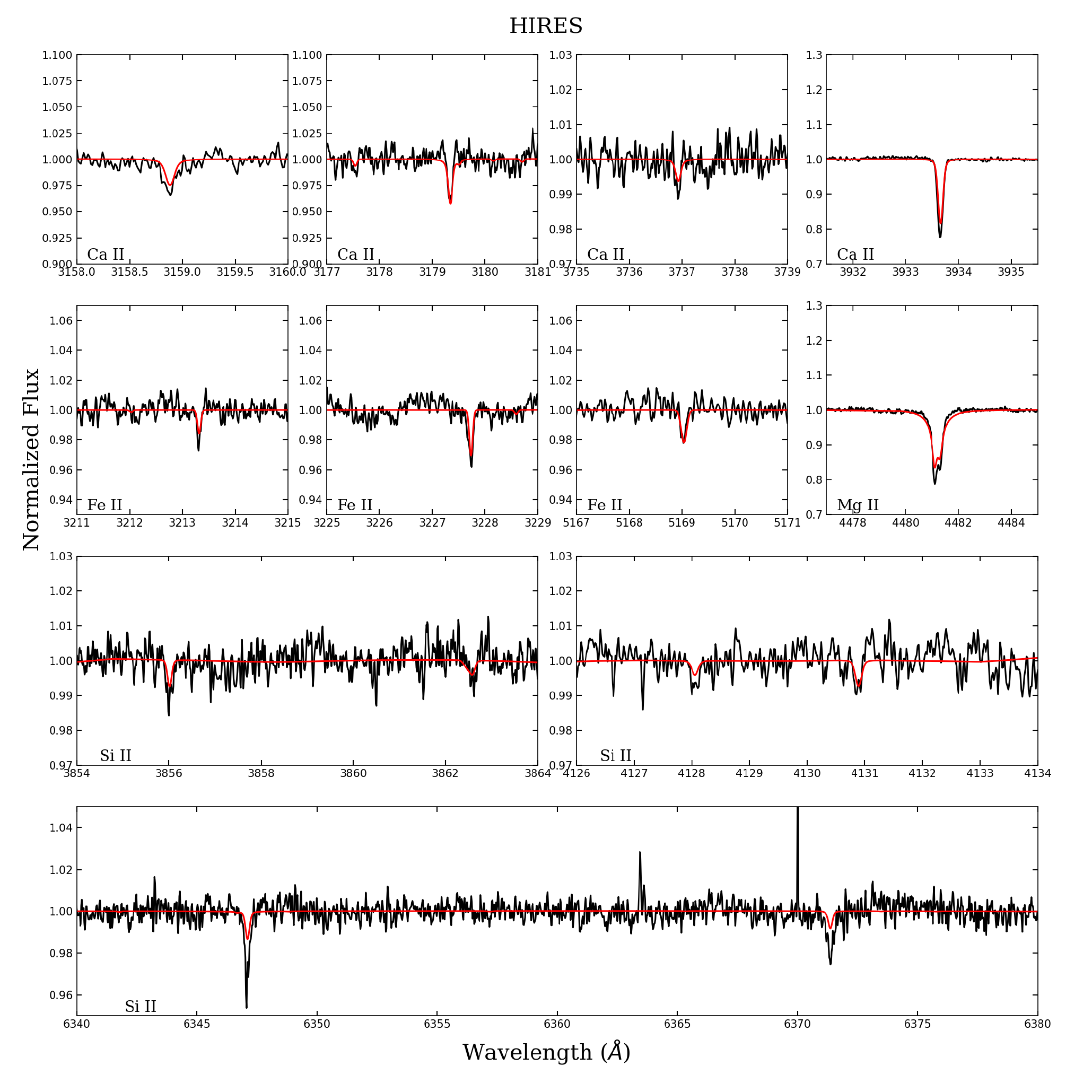}{6.5in}{}
\caption{HIRES spectra showing all detected lines (except Ca\,II $\lambda$3968 which lies within the
H$\epsilon$ Balmer transition). The model with the best-fit abundances is overplotted in red.}
\label{fig:hires}
\end{figure*}

\subsection{Abundances \label{sec:abn}}


For each absorption line we calculate an elemental abundance by linearly scaling a model input abundance by the ratio of EW measured in the observed spectrum to that of the model.   From the set of lines for each element we take the average abundance 
(giving all lines equal weight) as the input abundance for the next model iteration.  This process is repeated until the average abundance for each element converges, resulting in our final best-fit model.  Abundances are reported in Table \ref{tab:material}. 


\begin{deluxetable*}{ccccccc}
\tablecolumns{7}
\tabletypesize{\small}
\tablecaption{Accreting Material \label{tab:material}}
\tablehead{\colhead{at. \#} & \colhead{Z} & \colhead{$n(Z)/n(H)$} & \colhead{log($n(Z)/n(H)$)} & \colhead{$f_{\rm Z}$ ($10^{6}$ g/s)} & \colhead{n(Z)/n(O)} & \colhead{\% by mass}  }
\startdata
6 & C & $\phantom{<} (1.02 \pm 0.27) \times 10 ^{-6} $ & $\phantom{<} -5.99 ^{+0.10}_{-0.13}$ & $\phantom{<}2.1 \pm 0.5$ & $\phantom{<}0.19 \pm 0.06$ & $\phantom{<}3.5 \pm 0.7$ \\
7 & N & $\phantom{<} (1.47 \pm 0.30) \times 10 ^{-7} $ & $\phantom{<} -6.83 ^{+0.08}_{-0.10}$ & $\phantom{<}0.54 \pm 0.11$ & $\phantom{<}0.043 \pm 0.013$ & $\phantom{<}0.91 \pm 0.17$ \\
8 & O & $\phantom{<} (2.73 \pm 0.58) \times 10 ^{-6} $ & $\phantom{<} -5.56 ^{+0.08}_{-0.10}$ & $\phantom{<}14.3 \pm 3.0$ & - & $\phantom{<}24.1 \pm 4.0$ \\
12 & Mg & $\phantom{<} (2.17 \pm 0.54) \times 10 ^{-6} $ & $\phantom{<} -5.66 ^{+0.10}_{-0.12}$ & $\phantom{<}8.4 \pm 2.1$ & $\phantom{<}0.39 \pm 0.12$ & $\phantom{<}14.1 \pm 2.7$ \\
13 & Al & $\phantom{<} (5.88 \pm 1.30) \times 10 ^{-8} $ & $\phantom{<} -7.23 ^{+0.09}_{-0.11}$ & $\phantom{<}0.27 \pm 0.06$ & $\phantom{<}0.0111 \pm 0.0034$ & $\phantom{<}0.452 \pm 0.086$ \\
14 & Si & $\phantom{<} (3.99 \pm 1.36) \times 10 ^{-7} $ & $\phantom{<} -6.40 ^{+0.13}_{-0.18}$ & $\phantom{<}2.09 \pm 0.71$ & $\phantom{<}0.083 \pm 0.033$ & $\phantom{<}3.52 \pm 0.82$ \\
15 & P & $\phantom{<} (7.43 \pm 2.85) \times 10 ^{-9} $ & $\phantom{<} -8.13 ^{+0.14}_{-0.21}$ & $\phantom{<}0.05 \pm 0.02$ & $\phantom{<}0.0018 \pm 0.0008$ & $\phantom{<}0.084 \pm 0.021$ \\
16 & S & $\phantom{<} (1.72 \pm 0.72) \times 10 ^{-7} $ & $\phantom{<} -6.76 ^{+0.15}_{-0.24}$ & $\phantom{<}1.44 \pm 0.60$ & $\phantom{<}0.050 \pm 0.023$ & $\phantom{<}2.42 \pm 0.64$ \\
20 & Ca & $\phantom{<} (1.78 \pm 0.39) \times 10 ^{-7} $ & $\phantom{<} -6.75 ^{+0.09}_{-0.11}$ & $\phantom{<}1.90 \pm 0.42$ & $\phantom{<}0.053 \pm 0.015$ & $\phantom{<}3.20 \pm 0.57$ \\
22 & Ti & $< (4.52 \phantom{\pm 0.00}) \times 10 ^{-8} $ & $< -7.34 \phantom{^{+0.00}_{-0.00}}$ & $<0.6 \phantom{\pm 0.00}$ & $<0.01 \phantom{\pm 0.00}$ & $<1.1 \phantom{\pm 0.00}$ \\
24 & Cr & $< (3.07 \phantom{\pm 0.00}) \times 10 ^{-8} $ & $< -7.51 \phantom{^{+0.00}_{-0.00}}$ & $<0.5 \phantom{\pm 0.00}$ & $<0.01 \phantom{\pm 0.00}$ & $<0.9 \phantom{\pm 0.00}$ \\
25 & Mn & $< (1.46 \phantom{\pm 0.00}) \times 10 ^{-8} $ & $< -7.84 \phantom{^{+0.00}_{-0.00}}$ & $<0.3 \phantom{\pm 0.00}$ & $<0.01 \phantom{\pm 0.00}$ & $<0.5 \phantom{\pm 0.00}$ \\
26 & Fe & $\phantom{<} (1.31 \pm 0.40) \times 10 ^{-6} $ & $\phantom{<} -5.88 ^{+0.12}_{-0.16}$ & $\phantom{<}26.6 \pm 8.2$ & $\phantom{<}0.53 \pm 0.19$ & $\phantom{<}44.8 \pm 4.7$ \\
28 & Ni & $< (2.18 \phantom{\pm 0.00}) \times 10 ^{-7} $ & $< -6.66 \phantom{^{+0.00}_{-0.00}}$ & $<5.1 \phantom{\pm 0.00}$ & $\phantom{<}0.03 \pm 0.01$\tablenotemark{$\dagger$} & $2.9 \pm 0.6$\tablenotemark{$\dagger$} \\
\enddata
\tablecomments{$n(Z)/n(H)$ are the observed atmospheric elemental abundances by number. $f_{Z}$ is the diffusion flux calculated at optical depth $\tau$ = 1 assuming steady state accretion. $n(Z)/n(O)$ are calculated from $f_{Z}$ and represent the ratios in the accreting material.  
The total diffusion flux of heavy elements though the atmosphere (and hence, total accretion rate) is $(57.7 \pm 9.1) \times 10^6$ g s$^{-1}$. } 
\tablenotetext{\dagger}{These values are set such that Ni/Fe is equal to that of a CI Chondrite.}
\end{deluxetable*}

To derive uncertainties we take into consideration 
the spread of abundances from lines of the same element (standard deviation of the mean, SDOM) and abundance changes from varying $T_{\rm eff}$/log\,$g$. In the case of elements with only one line detected there is no SDOM; instead the propagated line measurement error is used.  It has been shown $-$ for helium-dominated WDs at least  \citep{2010ApJ...709..950K, 2011ApJ...741...64K} $-$ that element-to-element abundance ratios are only weakly dependent on changes in atmospheric parameters $T_{\rm eff}$/log\,$g$. Similarly, for G238-44 we find relative abundances are much less sensitive to variations in the stellar parameters than are absolute abundances.  
Fe, Si, and Mg display absorption lines in both UV and optical spectra for G238-44.  We show in Table \ref{tab:litcomp} the abundances calculated separately for the different instruments (and hence wavelength regimes).
Mg abundances are in excellent agreement, but there are discrepancies of $\simeq$0.2-0.3 dex in the nominal abundances of Si and Fe derived from UV and optical lines, although they do agree within the uncertainties. We also see a trend with Si ionization state: log(Si/H) $=$ $-6.36^{+0.13}_{-0.18}$, $-6.66^{+0.05}_{-0.06}$, $-7.04^{+0.08}_{-0.10}$ from Si II, Si III and Si IV, respectively.  
We ran tests to see if different $T_{\rm eff}$ models 
could bring the abundances for these different ionization states into better agreement, but found only minimal improvement. 
UV-optical discrepancies have been noted before \citep[e.g.][]{2012ApJ...750...69J, 2012MNRAS.424..333G, 2019AJ....158..242X, 2019MNRAS.483.2941W}.  \citet{2012MNRAS.424..333G} provided an in-depth consideration of possible origins of the UV-optical discrepancies they found in warm H-dominated WDs similar to G238-44 including uncertain atomic data, abundance stratification, and genuine variation of accretion rates, but no clear culprit was identified. Section \ref{subsec:absorp_lines} suggests there are no obvious variations in the accretion rate for G238-44.

\begin{deluxetable}{c|ccc}
\tablecolumns{4}
\tabletypesize{\normalsize}
\caption{UV versus Optical Abundances\label{tab:litcomp}}
\tablehead{ \colhead{} &
            \colhead{FUSE} &
            \colhead{STIS} &
            \colhead{HIRES} \\*[-3mm]
            \colhead{} &
            \colhead{905-1195 \r{A}} &
            \colhead{1150-1750 \r{A}} &
            \colhead{3000-10000 \r{A}}
}            
\startdata
Z & \multicolumn{3}{c}{log(Z/H)} \\*[0.5mm]
\hline
\\*[-2.5mm]
Mg  & $-$ & $-5.66 ^{+0.10}_{-0.12}$ & $-5.68 ^{+0.10}_{-0.12}$ \\*[2.0mm]
Si & $-6.66 ^{+0.08}_{-0.10}$ & $-$ & $-6.36 ^{+0.17}_{-0.29}$ \\*[2.0mm]
Fe  & $-6.06 ^{+0.10}_{-0.13}$  & $-5.76 ^{+0.12}_{-0.16}$ & $-5.83 ^{+0.12}_{-0.16}$ \\*[1.0mm]
\enddata
\end{deluxetable}

\subsection{Individual Elements \label{subsec:indiv_elems}}
Figures \ref{fig:fuse}$-$\ref{fig:c_multipanel} show representative elemental absorption lines. More detailed discussion for each element appears below.

\subsubsection{Carbon \label{subsubsec:C}}
\begin{figure}
    \centering
    \includegraphics[width=7cm]{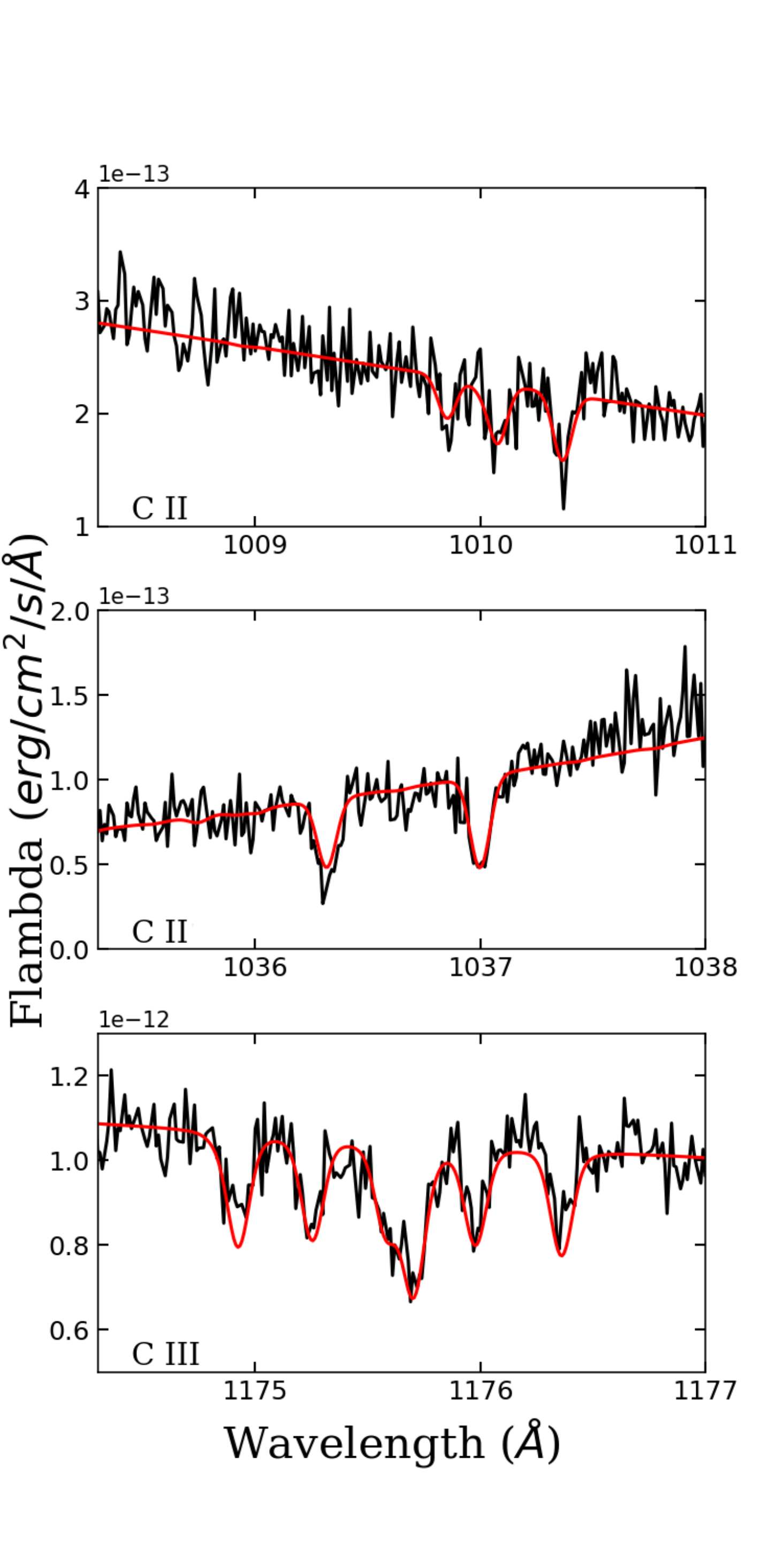}
    \caption{Carbon lines found in FUSE spectra. See Section \ref{subsubsec:C}}.
    \label{fig:c_multipanel}
\end{figure}

Absorption lines from C II and C III are present in FUSE spectra. The lines at 1036 and 1037\,\r{A} come from the ground state and a slightly excited state, respectively. All other C lines are from excited states and are photospheric. Figure \ref{fig:c_multipanel} shows that the 1036 and 1037\,\r{A} lines are consistent with a model of purely photospheric absorption.  Thus, any ISM component to their line strengths would have to be negligibly small and at the same radial velocity as the photospheric lines.

\subsubsection{Nitrogen \label{subsubsec:N}}
All four N lines detected in the FUSE spectrum come from the either the ground state or only slightly excited states (Table \ref{tab:linelist}), thus it is important to assess if they could originate in the ISM. In Table \ref{tab:nIIism} we show a comparison of the relative N~II line strengths for G238-44 to those of measured ISM lines toward the B1V star $\alpha$ Virginis located 77\,pc from Earth \citep{1979ApJ...228..127Y}.  The $\alpha$ Virginis measurements show that the ISM contribution to absorption in the excited states is significantly less, or absent, compared to the ground state line. Our measured relative line strengths for all four N~II lines observed in G238-44 are not consistent with the ISM measurements for $\alpha$ Virginis.  Given the relative observed line strengths, and agreement with predicted model photospheric abundances for all N lines in G238-44, we conclude that the nitrogen detected in its spectrum is photospheric.

\begin{center}
\begin{table}[h]
\caption{N\,II EWs for $\alpha$ Virginis and G238-44 \label{tab:nIIism} }
\begin{tabular}{*{5}{c}}
\hline 
\hline
$\lambda$ (\r{A})  & \multicolumn{2}{c}{Eq. Width (m\r{A})}  & $\chi$ (eV) & log($gf$) \\
        &   $\alpha$ Virginis & G238-44 \\
\hline
	1083.990	&	67.3	  & 47.5 & 0.000 & -0.939	\\
	1084.580	&	9.9	  & 60.8 & 0.006 & -0.587	\\
	1085.546	&		  & 40.2 & 0.016 & -1.079	\\
	1085.701	&	$<0.2$	  & 55.7 & 0.016 & -0.320	\\
\hline
\multicolumn{5}{p{8cm}}{Note. -- N\,II absorption in $\alpha$ Virginis is from ISM material \citep{1979ApJ...228..127Y}.}
\end{tabular}
\end{table}
\end{center}

\subsubsection{Oxygen \label{subsubsec:O}}
The O I 1152\,\r{A} line is detected in the FUSE spectrum. It is a clear, strong, line that comes from an excited state and is not expected to contain any contribution from the ISM.

\subsubsection{Magnesium \label{subsubsec:Mg}}
Mg is detected in the optical (blended) triplet at 4481\,\r{A}, arising from an excited state of 8.9\,eV. Four Mg~II lines are detected in the near UV around 2800\,\r{A}; two of these come from the ground state and two arise from excited levels of 4.4\,eV. 
The excited state lines must be photospheric in origin, and our model provides consistent fits to both optical and UV excited lines.   
Figure \ref{fig:fuse} shows that the model based on photospheric absorption fits all four near UV lines well, including the ground state lines previously attributed to the ISM by \citet{2014ApJ...787...75M}. We conclude that the ground state Mg~II lines are almost completely, if not entirely, also photospheric in origin.

\subsubsection{Aluminum \label{subsubsec:Al}}
Six Al lines are found in the HST/COS data. Three of these arise out of the ground state, but the other three are from an excited state of 4.6\,eV, and are thus photospheric. Our model fits to the photospheric lines do not suggest an ISM contribution to the ground state lines.

\subsubsection{Silicon \label{subsubsec:Si}}
Si II, III, and IV are found in FUSE data and Si II lines are found in optical data. We identified a total of 16 transitions; essentially all of these lines come from excited states and are necessarily photospheric.  
Comparison of abundances derived from the sets of UV and optical Si lines indicates discrepancies that have been previously noted for polluted WDs. See Section \ref{sec:abn} for more discussion.


\subsubsection{Phosphorus \label{subsubsec:P}}
The only phosphorus line observed is P II at 1153.995\,\r{A}; we regard it as a marginal detection.  This transition is slightly excited. \citet{1979ApJ...228..127Y} detect toward $\alpha$ Virginis 
the ground state P II line at 1152.818\,\r{A} from the ISM, but did not detect the 1153.995\,\r{A} line. Their study, with a resolution of 0.05\,\r{A}, had the resolving power to distinguish between the two lines. We do not detect P II absorption at 1152.818\,\r{A}, therefore any absorption from the excited state P II line at 1153.995\,\r{A} is not interstellar in origin.  The potentially stronger P II line at 1015.46\,\r{A} seen in other polluted WDs \citep[e.g.~GD 378,][]{2021ApJ...914...61K} is not detected in G238-44, which is unsurprising as it lies on the blue wing of Lyman-$\beta$ where the SNR is only 7, as compared to a SNR of 25 near 1153.995\,\r{A}. 

\subsubsection{Sulfur \label{subsubsec:S}}
Four S II lines are detected in FUSE data. The lines come from excited states and are not from the ISM.

\subsubsection{Calcium \label{subsubsec:Ca}}
Ca II is detected via 5 optical transitions. The weakest line detected in this study is the Ca II 3736\,\r{A} line with an equivalent width of 0.89 $\pm$ 0.27\,m\r{A}. This is made possible by the high SNR of these data and high resolving power of HIRES.


\subsubsection{Iron \label{subsubsec:Fe}}
Many Fe II lines are detected in FUSE, STIS, and HIRES data; Fe III is also seen in the FUSE spectrum.   Similar to the situation with Mg II STIS lines (Section \ref{subsubsec:Mg}), the ground state Fe II lines are well fit by our model at the derived photospheric abundance and do not reveal any significant contribution from the ISM.


\subsubsection{Titanium, Chromium, Manganese, and Nickel \label{subsubsec:upperlims}}
We measure upper limits for Ti, Cr, Mn, and Ni. These were determined by adjusting the abundance of these elements in the model until the absorption was comparable to the noise in their region of the HIRES spectrum.



\section{Discussion \label{sec:parent} }

Metals in the atmosphere of G238-44 should settle out of view within timescales on the 
order of days \citep{2009A&A...498..517K}. Combined with an essentially stable atmospheric
pollution level detected over decades, we conclude that G238-44 is in a steady state
accretion phase (meaning the inflow of material is balanced by losses due to settling). 
This is assumed to be the case for all further analysis wherein atmospheric
abundances are mapped to parent body compositions. Following \citet[][]{2012MNRAS.424..333G} and \citet{2014A&A...566A..34K}, we compute the diffusion flux
of metals through the visible atmosphere of G238-44 to arrive at the chemical composition of the accreting material (column {5} of Table \ref{tab:material}).






\subsection{Minerals, Metals, and Ices\label{subsec:o_budget}}

G238-44 provides detections {for many elements. Ni is an important element but not detected; we assume it is present in chondritic ratios to Fe as they have similar behaviors based on condensation temperatures and tendencies to be in a metallic form.}

Referring to the \% mass composition column in Table \ref{tab:material}, Fe is the most abundant element in the accreting material comprising nearly half of the total mass. This is already suggestive of a contribution from Fe in metallic form, with a mass percentage greater than that in bulk Earth \citep[29\%,][]{2001E&PSL.185...49A} though not so high as in planet Mercury \citep[69\%,][]{2018EPSC...12..404B}.

This stands in contrast to the substantial abundance of the volatile element N at 0.9\% by mass. Such a value is not at all characteristic of rocky planetary bodies that formed interior to the ice line of their host star; e.g.~bulk Earth has N by mass of roughly 0.00013\% \citep{2001E&PSL.185...49A}.  Instead, the N abundance found in the material polluting G238-44 is similar to outer solar system objects such as comet Halley \citep[1.5\%,][]{1988Natur.332..691J} and the exo-Kuiper Belt analog polluting G200-39 \citep[2\%,][]{2017ApJ...836L...7X}.  Such planetary bodies formed well beyond the ice lines in their protoplanetary discs where the stability of N$_2$ and ammonia ices allow these compounds to remain incorporated in solids as observed on the surfaces of Pluto and Charon  \citep{2015Icar..246...82C}.  It follows that a substantial amount of H$_2$O ice is also expected to go along with icy compounds of N and other volatiles such as C and S.

{Additionally, it can be seen in Table \ref{tab:material} that the Mg abundance is much higher than Si ($\sim$4.7 times by number).
In solar system rocks Mg/Si is usually $\sim$1 \citep[e.g.][]{2021SSRv..217...44L}.
We checked this ratio against abundance data from the Hypatia Catalog \citep{2014AJ....148...54H} and found it unlikely that such a ratio could be explained by nucleosynthetic variability \citep[e.g.][]{2010ApJ...715.1050B,2010ApJ...725.2349D,2021MNRAS.503.1877B}.
The observed abundance is unusual, but it does not change the interpretation given in Section \ref{subsec:twobody}. If Mg/Si were lower (i.e. ``normal''), then it would be even more difficult to describe the data with a single object.

}




To better assess the mineralogy of the material accreting onto G238-44 we calculate the oxygen budget following \citet{2010ApJ...709..950K}. This provides an accounting for accreted O 
under an initial assumption that all O is tied up in solid planetary system material as rocky
mineral oxides (e.g., MgO, Al$_2$O$_3$, SiO$_2$, P$_2$O$_5$, CaO, FeO or Fe$_2$O$_3$, and NiO;
in general molecules of the form ${\rm Z}_{p({\rm Z})}{\rm O}_{q({\rm Z})}$).
The oxygen budget O$_{budget}$ is defined as:

\begin{equation}
    {\rm O}_{budget} = \sum _{\rm Z} \frac{q({\rm Z})}{p({\rm Z})} \frac{n({\rm Z})}{n({\rm O})}
    \label{eq:oexcess}
\end{equation}

\noindent For a ``perfect'' rock ${\rm O}_{budget}$ $=$ 1, and in the case of a dry rocky parent body
this should be realized when considering only mineral oxides with the major and minor rock-forming elements.
If one initially obtains an O$_{budget}$ $<$ 1 when considering only mineral oxides, then there is an oxygen excess and the additional O atoms are 
presumed to be associated with ices (H$_2$O and possibly CO and CO$_2$ if sufficient C is present).
If an O$_{budget}$ $>$ 1 is initially obtained then there is an oxygen deficit and it is assumed that
metallic material (typically Fe) is present as is found within the metal cores of differentiated bodies in our solar system.


Assessing O$_{budget}$ for G238-44 using the values given in Table \ref{tab:material} and only the
rocky mineral oxides listed above (i.e., no ices or consideration of C or N), we find it to be 
nearly balanced within the various abundance uncertainties (${\rm O}_{budget}$ $\simeq$ 1).
However, this leaves the significant contribution from C and N to the parent body 
($\approx$4.4\% by mass) unaccounted for. Carbon and nitrogen can be locked within organic materials 
that are found in trace amounts in solar system bodies, and indeed are the most abundant C- and N-bearing phase within
carbonaceous chondrites (e.g., \citealt{pearson06}). The C and N in G238-44
could be present as aromatic organic compounds composed only of C, N, and H (e.g., Pyrrole, Aniline, Pyrazine, Purine, etc.). However, such a parent body would be very unusual, effectively a rock
marbled with organic substances. Lacking a reasonable analog to such a parent body within the solar
system, we opt instead for the C and N to be in the more familiar form of ices.

As noted earlier, the presence of C- and N-bearing ices heralds the presence of icy H$_2$O.
H$_2$O, CO, and CO$_2$ ices will ``steal'' O away from rocky minerals thus leading to deficit
conditions, meaning the inclusion of ices can only be balanced by allowing for metallic material.
Again, a single body hosting significant quantities of icy material $-$ especially C- and N-bearing
ices $-$ and metallic Fe would be unlike anything currently known within the solar system. {Even the EH Chondrites -- primitive rocks rich in Fe, candidates for the composition of an early Earth \citep{2010E&PSL.293..259J} -- are a poor description of the data.}

 
To explain what appears to be the simultaneous existence of significant icy and metallic portions in the material being accreted by G238-44 we propose the presence of two distinct parent bodies, one core-like and one volatile-rich. Below we motivate such a scenario with the help of literature studies
and explore what two parent bodies would be capable of producing the observed abundances in G238-44.

\subsection{Simultaneous Accretion From Two Parent Bodies \label{subsec:twobody}}

\begin{figure*}
    \centering
    \includegraphics[width=0.9\textwidth]{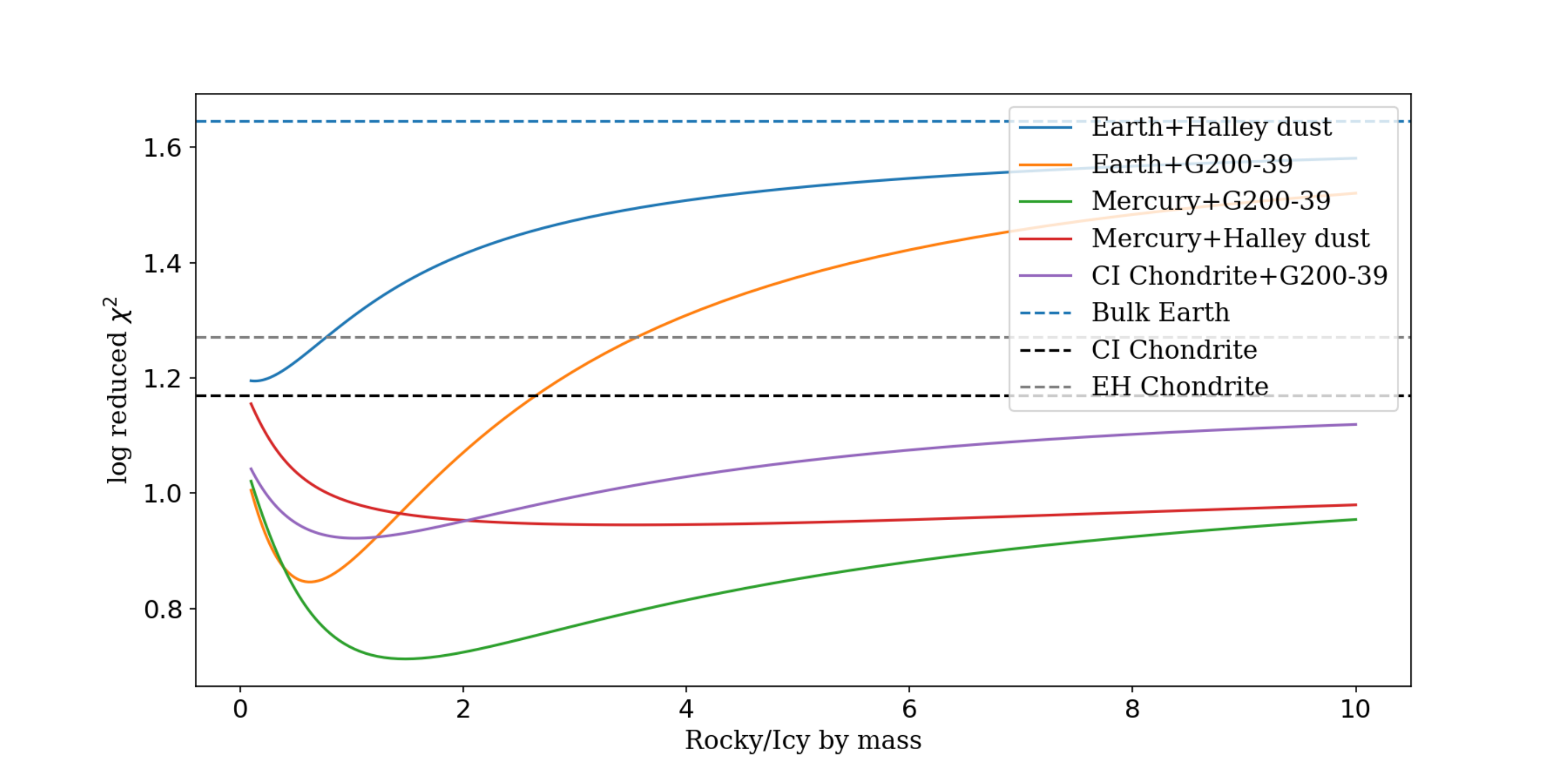}
    \caption{Reduced $\chi^2$ fits exploring two-body models, note that the y-axis is log$_{10}$. 
    Only atomic abundance information is incorporated into these model tests; i.e., the presence
    of metallic and icy material is not assessed by the $\chi^2$ fit.
    We find that a mix of metal-dominated Mercury-like material and icy Kuiper Belt object-like material (G200-39; see text) in a ratio of 1.7:1 provides the best fit to the data. 
    }
    \label{fig:chi2}
\end{figure*}

\begin{figure*}
    \centering
    \includegraphics[width=0.9\textwidth]{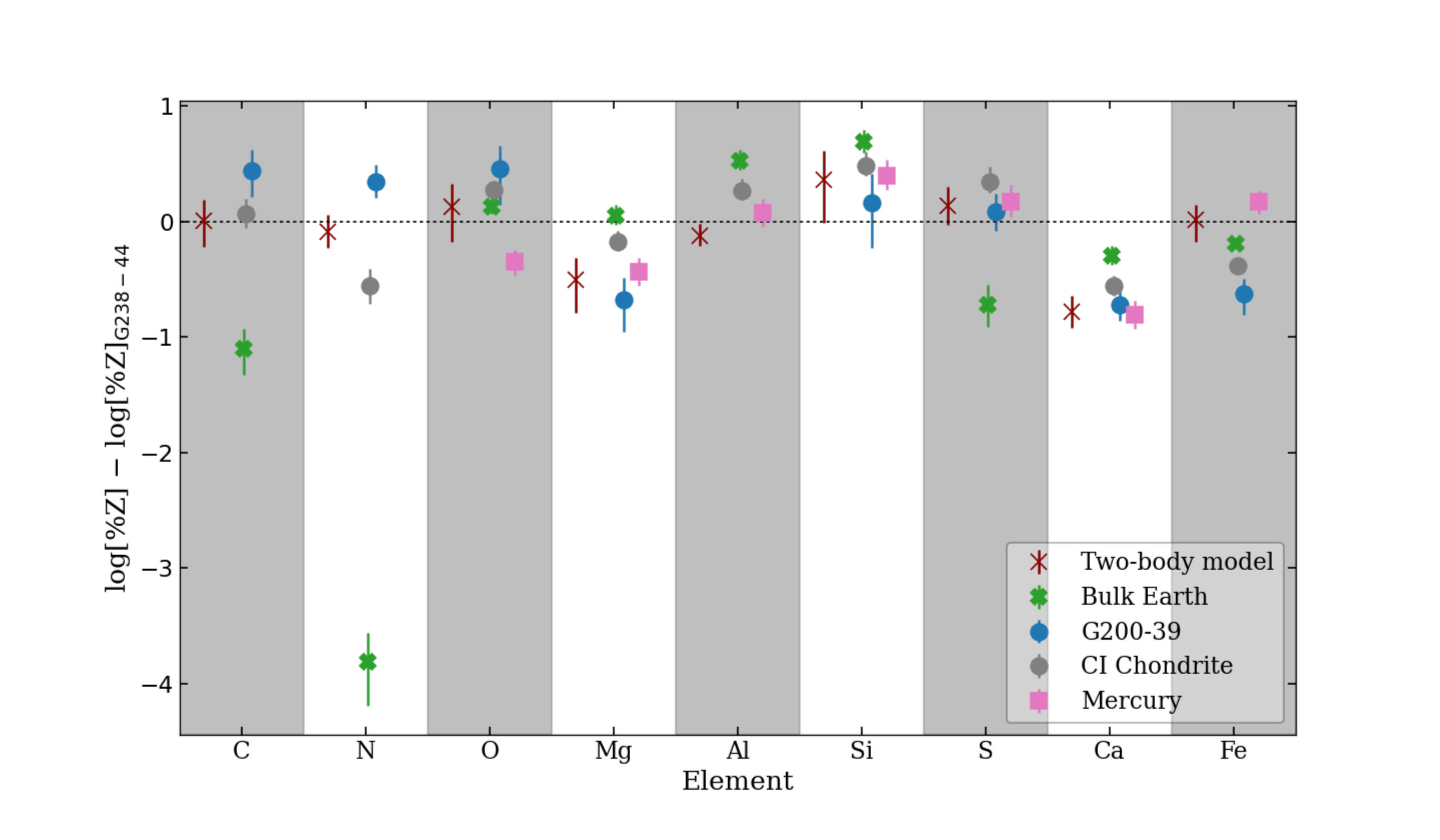}
    \caption{Residuals for elemental abundances (expressed as a percentage of the total parent body mass) when comparing G238-44 and various parent body models. The two-body model, informed from the best fit in Figure \ref{fig:chi2} and discussed in Section \ref{subsec:twobody}, fits within the errors both for Fe and ices (C, N, O). Single-body models are not reasonable matches to the data.}
    \label{fig:residuals}
\end{figure*}

\begin{figure*}
    \centering
    \includegraphics[clip, trim=0.1cm 0.1cm 10.5cm  0.1cm, width=\textwidth]{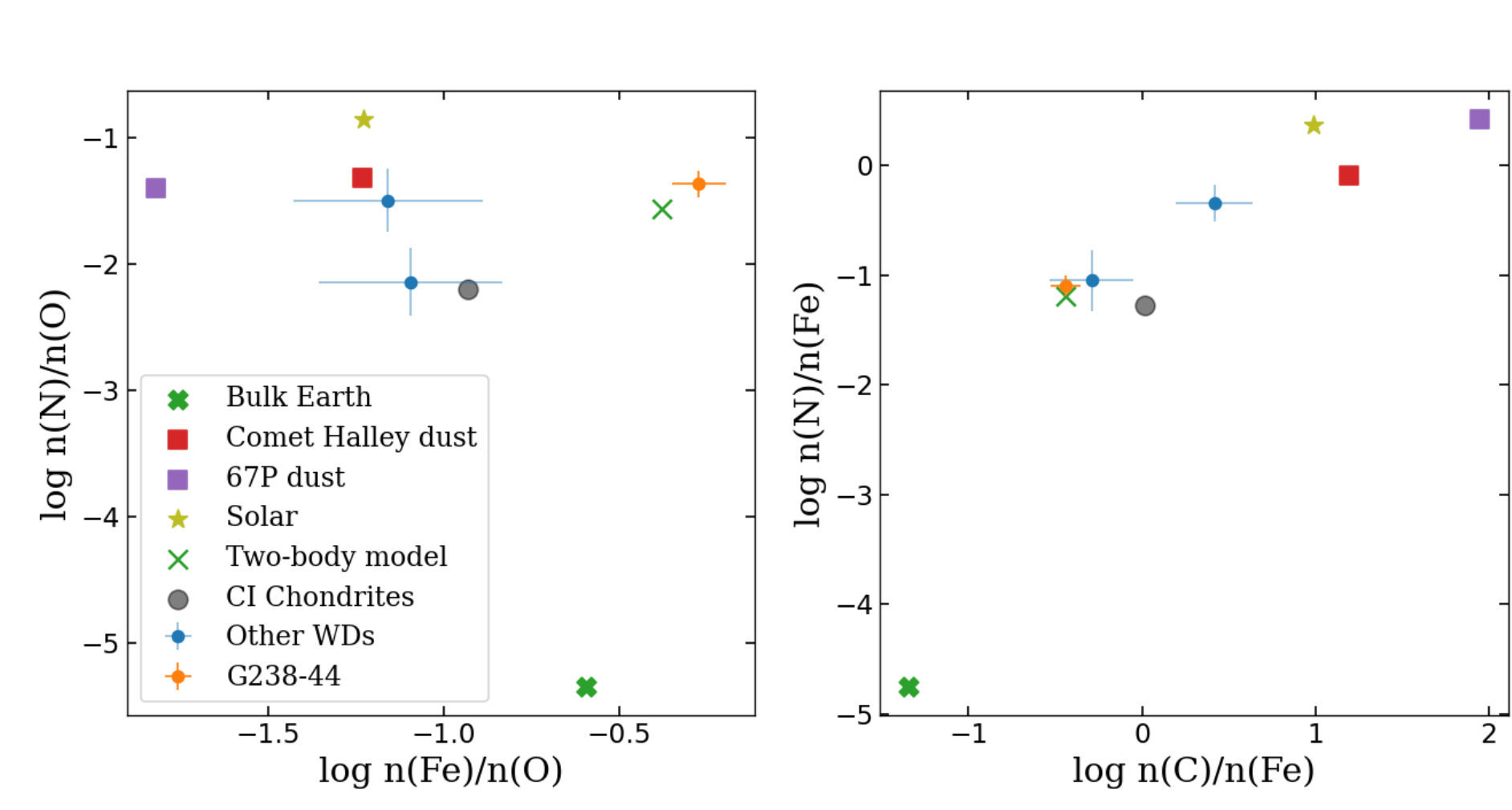}
    \caption{Elemental abundance ratios by number for the material polluting G238-44 compared to other WDs, solar system bodies, and the two-body model we propose in Section \ref{subsec:twobody}.
    Abundances from: Bulk Earth -- \citet{2001E&PSL.185...49A}, Mercury -- \citet{2018EPSC...12..404B,2018mvam.book...30N}, Comet Halley dust -- \citet{1988Natur.332..691J}, 67P/Churyumov–Gerasimenko -- \citet{2017MNRAS.469S.712B}, Solar \& CI Chondrite -- \citet{2021SSRv..217...44L}, G200-39 -- \citet{2017ApJ...836L...7X}, GD 378 -- \citet{2021ApJ...914...61K}. 
    O is an incredibly versatile element when it comes to planet formation. It is a significant component of any solar system object. C, N, and Fe however are much more specialized. Fe abundance is high in objects that formed close to their parent star. Contrary, N can only form in significant amounts past the ice line. We do not expect objects that are high in Fe to also be high in N. G238-44 breaks this trend and has both high Fe/O and N/O. The proposed two-body model is
    capable of reproducing this unusual characteristic.
    }
    \label{fig:ZtoMg}
\end{figure*}

The material accreting onto G238-44 has two opposing characteristics:

\begin{itemize}
    \item {\bf Significant Portion of Iron in Metallic Form}. Fe metal indicates a core-like structure of a differentiated body, likely forming near its host star as an asteroid or terrestrial planet \citep{2019JTecto.760.165T}.
    \item {\bf High Concentration of Nitrogen}. A large N abundance indicates formation far from the host star, well beyond the ice line, as in Kuiper Belt objects \citep{2019Icar..326...10B}. Such minor planets are rich in ices of N, C, and O.
\end{itemize}

\noindent The characteristics of the material polluting the atmosphere of G238-44 suggest that it originated from two distinct formation locations of the host stars' planetary system. We invoke a model wherein there
are two compositionally distinct parent bodies simultaneously polluting G238-44, one coming from
the inner planetary system (e.g., an asteroid-like body) and the other coming from the outer planetary
system (e.g., a Kuiper Belt object analog).

Literature works have laid the foundation for such an interpretation, suggesting that not only is it
possible to have two bodies simultaneously polluting a white dwarf atmosphere but that for smaller
bodies it may be common. \citet{2008AJ....135.1785J} first proposed multiple bodies being captured
around a white dwarf to explain ``invisible'' accretion reservoirs (like the one around G238-44).
In their model, a second body approaching a WD has a trajectory that is inclined to an existing accretion disk of an earlier tidally-disrupted body which results in an impact that causes grains from both objects to vaporize due to sputtering. The disk is then mainly composed of gas with an infrared excess that would be difficult to detect.

Recent works have consistently indicated that a continuous stream of smaller bodies should
be arriving at the white dwarf star (e.g., \citealt{wyatt14,2020MNRAS.491.4672T,li22,trierweiler22}, and references
therein). The work of \citet{li22} is especially relevant in this context as it explores
(for the solar system) the transport of main-belt asteroids and trans-Neptunian objects (or Kuiper Belt 
objects) to a white dwarf as a function of time. Specifically, they find it is possible to deliver
both a main-belt asteroid and Kuiper Belt object to the central white dwarf star within a white dwarf
cooling time of 100\,Myr. Events involving an inner planetary system body and outer planetary
system body are rare, though, as Kuiper Belt object pollution is supposed to 
increase in frequency for cooling times $\gtrsim$100\,Myr \citep{li22}. 
This would be consistent with G238-44
being the first white dwarf we know of to possibly be experiencing a simultaneous pollution event
of such type.


The results from theoretical and simulation work to date support the possibility of
having two bodies from distinct formation locations around G238-44 now simultaneously
polluting the white dwarf star. Next we must verify that a two-body model is capable of
producing a reasonable fit to the abundance data.

In arriving at a two-body fit to the data we envision a metallic 
parent body with mass comparable to an asteroid but with composition 
similar to Mercury {\citep{2018EPSC...12..404B,2018mvam.book...30N}}. This choice is made as Mercury is the 
best-characterized metal-dominated body we know of in the solar system. 
The asteroid 16 Psyche may be a more appropriate analogy based on the fact
that it is the largest M-type (metal-rich) asteroid known, but it lacks 
detailed elemental composition estimates (the Psyche asteroid mission will launch {in the future} to better characterize this interesting object). As a model for the icy 
body we adopt the observed abundances of the extrasolar Kuiper Belt object-analog G200-39 \citep{2017ApJ...836L...7X}. This measurement is more relevant than in situ measurements of comets because it fully captures the bulk composition, which in the case of a Kuiper Belt object is likely unchanged since its formation. Recent measurements of comet dust \citep[e.g.][]{1988Natur.332..691J,2017MNRAS.469S.712B} do not account for the entire body, and short period comets have had their volatile compositions altered significantly by their multiple passes close to the Sun. Pluto, which is expected to be a core made up of hydrated rock covered by an ice-rich mantle and crust \citep{2017Icar..287....2M}, could be a relevant icy body analog but is not characterized well
at an elemental level.

In utilizing the parent body polluting G200-39 as a reference for our icy body abundances we
revisited its water-ice fraction. 
Accounting for CO as well as mineral oxides, we arrive at a somewhat different result than quoted in \citet{2017ApJ...836L...7X}.  We find that 65\% of the O is excess, corresponding to a water fraction of 46\% by mass.  By number, the ratio of excess O atoms to N is 19.7, and the ratio of excess O to C is 3.4.  These numbers are roughly consistent with measurements of solar system comets \citep[][]{2011ARA&A..49..471M}.

Figures \ref{fig:chi2} and \ref{fig:residuals} explore $\chi^2$ minimization for a variety of different two-body models
(and reference single-body models). We allowed the relative mass fraction of rocky-metal to
icy material to vary and fit against all measured elemental abundances. The Mercury+G200-39
model significantly outperforms all other models, especially for a 1.7:1 mix of Mercury-like and
icy Kuiper Belt object-like material.

Figure \ref{fig:ZtoMg} shows how this particular model produces C/N/O/Fe abundance ratios that
are in good agreement with measurements for G238-44 and yet distinct from other solar system
bodies.



We are thus able to identify a mix of two parent bodies that reproduces the observed pollution
pattern in the atmosphere  of G238-44. Based on the elemental diffusion fluxes in column 6
of Table \ref{tab:material}, we can make an estimate of how big each of these distinct parent bodies
may have been. 

In aggregate, G238-44 hosts
a total mass accretion rate of $5.8 \times 10^{7}$\,g\,s$^{-1}$. The star has been known to have pollution in its atmosphere since its first high-dispersion IUE observation in 1982. 
Assuming constant accretion over the past 40 
years, the star has swallowed $7 \times 10^{16}$\,g.  
However, we don't know how long it has been since the currently observed accretion began
nor how long it will continue forward. We can get an idea of a range of the total mass of the parent bodies using previous estimates for WD disk lifetimes of between 0.03 and 5 Myr \citep{2012ApJ...749..154G}.  Taking the simplifying assumption that the observed accretion rate is representative of a time-averaged mean accretion rate over the disk lifetime,
we arrive at a total pollution mass roughly between $6 \times 10^{19}$ and $9 \times 10^{21}$\,g. 
Taking now the 1.7:1 mass ratio of rocky-metal to icy material, we arrive at a metal-rich body
mass of between 3.8 $\times$ 10$^{19}$ and 5.7 $\times$ 10$^{21}$\,g and an icy body mass of between 2.2 $\times$ 10$^{19}$ and 3.3 $\times$ 10$^{21}$\,g. For reference, the mass of asteroid 16 Psyche is
roughly 2.3 $\times$ 10$^{22}$\,g and the mass of the primary within the trans-Neptunian system 
47171 Lempo is roughly 6.7 $\times$ 10$^{21}$\,g. The two parent bodies possibly polluting G238-44 would
each be roughly an order of magnitude less massive than these solar system objects.

\section{Conclusions \label{sec:conclusions}}
UV and optical spectroscopic data are used to probe heavy elements in the atmosphere of the warm DA white dwarf G238-44. HIRES observations over a baseline of 24 years show no robust evidence for variation in the equivalent width of optical metal absorption lines. With a diffusion timescale of days, we find the accretion rate onto the surface of G238-44 does not vary on a timescale of years. The temperature and brightness of this star make it an optimal candidate for further monitoring to detect any change in accretion rates.



Modeling of the spectra allows us to determine the abundances of heavy elements in the star's atmosphere, from which we infer the composition of the material
that pollutes it.
Within the uncertainties we are able to determine that the parent material is rich in nitrogen and likely contains a significant amount of metallic iron. This mix is unlike any known solar system body. We suggest that G238-44 is simultaneously accreting a metal-rich exo-planetesimal and a volatile rich exo-Kuiper Belt object. If our interpretation is correct, this would be the first evidence of simultaneous accretion of two distinct parent bodies in a white dwarf.  

The proposed two-body pollution interpretation may explain G238-44's lack of infrared excess \citep{2008AJ....135.1785J}. While such events are predicted in dynamical simulation work
(e.g., \citealt{li22}), they should be very rare for a white dwarf with cooling age $<$100\,Myr
like G238-44. This could be suggestive of a planetary system architecture for G238-44 different from
the solar system, the likes of which 
allowed the rapid and dramatic perturbation of inner and outer planetesimal populations
around G238-44.




\acknowledgements
We are grateful to K.\ Long and B.\ G\"{a}nsicke for providing their reduced and combined version of the FUSE spectrum. 
We are grateful for helpful advice and conversations with Alexandra Doyle, Isabella Trierweiler{, and Edward Young}.

This work has been supported by grants from NASA and the NSF to UCLA.  T.J.\ acknowledges support from the UCLA Physics and Astronomy summer 2020 REU program and 2021-22 URSP scholarship.
B.K.\ acknowledges support from the NASA Exoplanets program award number 80NSSC20K0270 (PI Ed D. Young).
C.M.\ and B.Z.\ acknowledge support from NSF grants SPG-1826583 and SPG-1826550.

This research has made use of the Keck Observatory Archive (KOA), which is operated by the W.M.\ Keck Observatory and the NASA Exoplanet Science Institute (NExScI).

This research has made use of SIMBAD, SAO/NASA ADS, the Montreal White Dwarf Database, MAST, VALD, NIST, and Kurucz Atomic Line databases, IRAF, Python, and Matplotlib.  



Some of the data presented in this work were obtained at the W.M.\ Keck Observatory, which is operated as a scientific partnership among the California Institute of Technology, the University of California and the National Aeronautics and Space Administration. The Observatory was made possible by the generous financial support of the W.M.\ Keck Foundation.  We recognize and acknowledge the very significant cultural role and reverence that the summit of Mauna Kea has always had within the indigenous Hawaiian community.  We are most fortunate to have the opportunity to conduct observations from this mountain. We thank the Keck Observatory staff for their dedicated work and support.

\facilities{FUSE, HST(STIS,COS), Keck~I(HIRES)}

\bibliographystyle{apj}
\bibliography{wd}

\end{document}